\documentclass{aa}  

\usepackage{graphicx}
\usepackage{txfonts}
\usepackage{placeins}
\begin{document}

   \title{Are Am stars and hot-Jupiter planets related?}


   \titlerunning{Am stars and hot-Jupiter planets}
   \authorrunning{Saffe et al.}

   \author{C. Saffe\inst{1,2,8}, J. Alacoria\inst{1,8}, P. Miquelarena\inst{1,2,8},
           R. Petrucci\inst{6,8}, M. Jaque Arancibia\inst{3,4}, R. Angeloni\inst{5}, E. Martioli\inst{7}, 
           M. Flores\inst{1,2,8}, E. Jofr\'e\inst{6,8}, A. Collado\inst{1,2,8} and F. Gunella\inst{1,8}
           }

\institute{Instituto de Ciencias Astron\'omicas, de la Tierra y del Espacio (ICATE-CONICET), C.C 467, 5400, San Juan, Argentina.
         \and Universidad Nacional de San Juan (UNSJ), Facultad de Ciencias Exactas, F\'isicas y Naturales (FCEFN), San Juan, Argentina.
         \and{Instituto de Investigaci\'on Multidisciplinar en Ciencia y Tecnolog\'ia, Universidad de La Serena, Ra\'ul Bitr\'an 1305, La Serena, Chile}
         \and Departamento de F\'isica y Astronom\'ia, Universidad de La Serena, Av. Cisternas 1200 N, La Serena, Chile.         
         \and Gemini Observatory / NSF’s NOIRLab, Casilla 603, La Serena, Chile
         \and Observatorio Astron\'omico de C\'ordoba (OAC), Laprida 854, X5000BGR, C\'ordoba, Argentina.
         \and Laborat\'orio Nacional de Astrof\'isica (LNA/MCTI), rua Estados Unidos 154, Itajub\'a, MG, Brasil
        \and Consejo Nacional de Investigaciones Cient\'ificas y T\'ecnicas (CONICET), Argentina
         }

   \date{Received xx, 2022; accepted xx, 2022}

 
  \abstract
   {Metallic-lined Am stars are often components of short-period binary systems, where tidal interactions would result in
   low rotational velocities and help to develop the chemical peculiarities observed.
   However, the origin of single Am stars and Am stars that belong to wide binary systems is unclear.}
   {There is very recent evidence of an Am star hosting a hot-brown dwarf likely synchronized and other possible Am stars hosting hot-Jupiter planets.
    Following literature suggestions, we wonder if these hot-low mass companions could play a role in the development of an Am star,
    that is to say, if they could help to mitigate the "single Am" problem.}
   {We carried out a detailed abundance determination via spectral synthesis of 19 early-type stars hosting
   hot-brown dwarfs and hot-Jupiter planets, in order to determine the possible presence of Am stars in this sample.
   The abundances were determined iteratively for 25 different species by fitting synthetic spectra using the SYNTHE program
   together with local thermodynamic equilibrium (LTE) ATLAS12 model atmospheres.
   The abundances of \ion{C}{I}, \ion{O}{I} and \ion{Mg}{I} were corrected by non-LTE effects.
   The complete chemical patterns of the stars were then compared to those of Am stars and other chemically peculiar stars.}
   {We studied a sample of 19 early-type stars, 7 of them hosting hot-brown dwarfs and 12 of them hosting hot-Jupiter planets.
  We detected 4 Am stars in our sample (KELT-19A, KELT-17, HATS-70 and TOI-503)
  and 2 possible Am stars (TOI-681 and HAT-P-69).
  In particular, we detected the new Am star HATS-70 which hosts a hot-brown dwarf, and rule out
  this class for the hot-Jupiter host WASP-189, both showing different composition than previously reported.
  For the first time, we estimated the incidence of Am stars within stars
  hosting hot-brown dwarfs (50-75\%) and within stars hosting hot-Jupiters (20-42\%).
  In particular, the incidence of Am stars hosting hot-brown dwarfs resulted higher than the frequency of
  Am stars in general.
  This would imply that the presence of hot-brown dwarfs could play a role in the development of Am stars
  and possibly help to mitigate the "single Am" problem, different to the case of hot-Jupiter planets. 
  Notably, these results would also indicate that the search for hot-brown dwarfs may be benefited by targeting single Am stars or Am stars in wide binary systems.
  We encourage the analysis off additional early-type stars hosting hot-companions in order to improve the significance of the 
  initial trends found here. 
  
  }
   {}


   \maketitle
%

\section{Introduction}

The metallic-lined A stars (Am stars) were first identified by \citet{titus-morgan40} as
objects with strong metal lines and a weak \ion{Ca}{II} K line in their spectra.
On average, Am stars present over-abundances of most heavy elements in their spectra, 
particularly Fe and Ni, together with under-abundances of Ca and Sc 
\citep{preston74,fossati07,fossati09,catanzaro19}. 
Starting with the work of \citet{conti70}, other authors prefer a wider definition for the group,
including stars with underabundances of Ca (and/or Sc) and/or overabundance of the Fe group elements
\citep[e.g. ][]{lane-lester87,adelman94,adelman97,yuce-adelman14}.
Am stars rotate more slowly than average A-type stars, as first noted by \citet{slettebak54}
and then confirmed by a number of works \citep[e.g. ][]{slettebak55,abt00,niemczura15}.
The origin of the peculiar abundances is commonly attributed to diffusion processes due to gravitational settling
and radiative levitation \citep{michaud70,michaud76,michaud83,vauclair78,alecian96,richer00,fossati07},
where stable atmospheres of slowly-rotating A-type stars would allow the diffusion processes to 
operate\footnote{Chemical peculiarities could also reach early F stars, the Fm stars.}.
Initial observations showed that Am stars do not pulsate \citep{breger70,breger72}, while further works
showed that indeed the pulsations are possible in a number of Am stars 
\citep[e.g. ][]{bessell-eggen72,kurtz89,smalley11,smalley17}.
This showed that pulsations do not necessarily conflict with the peculiar abundances 
observed in the Am stars.

A number of works show that Am stars are often components of binary systems \citep{abt61,abt-levy85,north98,debernardi00,cp07,stateva12a}.
In fact, the slow rotation of Am stars is usually attributed to a tidal braking
in short-period (1 d $<$ P $<$ 10 d) binary systems \citep{abt67,budaj96,budaj97},
rather than to the presence of weak or ultra-weak magnetic fields \citep[e.g., ][]{folsom13,blazere16,blazere20}.
Nonetheless, there are Am stars in wide binary systems with longer periods and Am stars that appear to be single
\citep[the "single Am" problem, ][]{murphy13}.
For instance, \citet{smalley14} suggest that 30\%-40\% of Am stars are either single or in wide binary systems,
while \citet{murphy12} presented the intriguing case of KIC 3429637, a single pulsating Am star observed by {\it{Kepler}}
with no evidence of frequency modulation by a possible companion.
In these cases of single Am stars or in wide binary systems, the origin of both Am stars and their
low rotational velocities is unclear.

Hot-Jupiter planets present short orbital periods {($<$ 10 d)} and large planetary masses {($>$ 0.1 M$_\mathrm{Jup}$)},
that is, they are gas giants orbiting very close to their stars \citep[e.g., ][]{wang15}.
By analogy, we refer to as hot-brown dwarfs (hot-BDs) those objects with short orbital periods and having masses 
above $\sim$13 M$_\mathrm{Jup}$, the minimum mass to burn Deuterium \citep[e.g., ][]{grossman73,saumon96}.
Recently, \citet{subjak20} (hereafter, S20) discovered TOI-503, the first Am star which hosts a hot-BD companion.
They found that the rotation period of the star is similar to the orbital period of the hot-BD (P$_{orb}\sim$ 3.67 d),
which could be indicative of synchronism between them.
They suggest that the presence of close low-mass companions to Am stars could be possibly invoked to explain their
small rotational velocities, where the tidal braking of a more distant stellar companion is not significant.
Then, \citet{addison21} (hereafter, A21) found an ultra-hot
Jupiter\footnote{Ultra-hot Jupiter planets present dayside temperatures of 2200 K or more \citep{cb19}.}
with a mass of 3.12 M$_\mathrm{Jup}$ orbiting around the star MASCARA-5/TOI-1431, which is found to be also a peculiar Am star.
They wonder if the presence of a giant planet can contribute to their host being an Am star, without evidence of a stellar binary.
In this way, both works S20 and A21, suggest a possible link between Am stars and the presence of
hot low-mass companions (hot-Jupiters and hot-BDs), with a lower influence of an additional stellar component.
Notably, it is also worthwhile to mention the recent detection of four additional early-type stars (KELT-17, KELT-19A, KELT-26 and WASP-189) hosting
hot-Jupiter companions, being confirmed or likely Am stars \citep{saffe20,siverd18,rodriguez20,lendl20}.

Can hot low-mass companions help to mitigate the "single Am" problem?
Can they facilitate -in some way- the rise of Am stars, as suggested by S20 and A21?
What is the frequency of Am stars (and chemically peculiar stars in general) hosting hot-low mass companions?
These important questions could be addressed searching for planets around Am stars, as suggested by A21.
Alternatively, it is also possible to perform a chemical analysis of early-type stars hosting hot-low mass companions
to determine the likely presence of Am stars within this group.
This is one of the motivations of the present work.
It would be easier to confirm systems with low $v\sin i$, 
which could preferentially be selecting stars which are susceptible to the Am phenomenon.
In the last few years, the population of hot-low mass companions around early-type stars is slowly growing compared to late-type stars
\citep[e.g. ][]{siverd18,rodriguez20,lendl20}, giving us the opportunity to perform this analysis.
If such possible relation between Am stars and hot-low mass companions is confirmed, specific surveys targeting
(single) Am stars could be important in helping to detect hot-Jupiters and/or hot-BDs. In addition, these ideas
could provide valuable insights into the formation and evolution of both Am stars and their companions.
 
Stars that apparently display a peculiar spectra could be confirmed (or rejected) through a detailed abundance analysis.
For example, \citet{catanzaro19} studied a group of 62 Am stars, previously selected with an abundance-based
criterion (subsolar values of Ca, Sc and overabundances of Ti, Cr, Mn and Fe).
For the case of stars with planets, the exoplanet host star WASP-33 was initially classified as Am \citep{grenier99}
and mentioned as such in different works (S20, A21).
However, \citet{collier10} caution that no obvious Am characteristics are visible in the spectrum of this star,
other than slightly weak \ion{Ca}{II} H\&K lines. Then, a recent abundance analysis including several species showed
that WASP-33 do not belong to this class \citep{saffe21}.
Conversely, the exoplanet host star KELT-17 was initially considered as a solar-like composition object, 
while a detailed chemical analysis showed a clear Am pattern \citep{saffe20,saffe21}.
These examples highlight the importance of a detailed chemical analysis including several species,
in order to confidently detect a bona fide Am chemical pattern.
This motivates us to perform a detailed chemical analysis of early-type stars hosting hot low-mass companions
in order to compare its chemical pattern with those of Am stars and other chemically peculiar stars.

Our sample includes a number of stars without previous detailed abundance analysis (such as TOI-587, TOI-681, CoRoT-3, KELT-1,
KELT-25, MASCARA-4, HAT-P-69 and HAT-P-70), being to our knowledge the largest group of 
early-type stars with hot-companions homogeneously studied to date.
We report the detection of the new Am star HATS-70 which hosts a hot-BD, and rule out the Am class for WASP-189
which hosts a hot-Jupiter planet (in both cases, our results differ from the composition previously reported).
This work is part of our ongoing program aimed to study early-type stars with planets \citep{saffe20,saffe21},
which are poorly studied in general compared to their late-type FGKM counterparts.

This work is organized as follows. In Sect. 2, we describe the observations and data reduction. In Sect. 3,
we present the stellar parameters and chemical abundance \mbox{analysis}. In Sect. 4, we show the results and
discussion. Finally, in Sect. 5, we highlight our main conclusions.


\section{Stellar samples and observations}

We start by compiling a list of early-type stars hosting hot-Jupiters and/or hot-BDs taken from the 
Extrasolar Planets Encyclopaedia\footnote{http://exoplanet.eu/}.
We also included a group of early-type stars with hot-BDs taken from the literature
\citep{deleuil08,collier10,siverd12,siverd18,bieryla14,lund17,talens17,gaudi17,temple17,
zhou16,zhou19a,zhou19b,subjak20,rodriguez20,lendl20,dorval20,grieves21}.
The complete sample studied in this work is shown in Table \ref{tab.sample}.
We present the star names, visual magnitude V, instrument, the signal-to-noise S/N measured near 5000 \AA,
together with information about their low-mass companions, that is, orbital period and mass.
The last column presents the reference for the orbital period and mass of their low-mass companions.
In this way, the final sample comprises 19 objects, including mainly A-type and some early F-type stars,
all having low-mass companions. Within this group, 7 stars host hot-BDs and 12 stars host hot-Jupiters.
Aiming to increase the number of stars, we included 8 stars with previously obtained stellar parameters \citep{saffe20,saffe21}.
This allows us to obtain, to our knowledge, the largest possible sample of early-type stars with 
hot-low mass companions analyzed to date, with uniformly derived abundances.
We caution that the sample includes also the star TOI-681 for completeness, with a somewhat larger BD orbital period
of 15.78 d \citep{grieves21}.

\begin{table*}
\centering
\caption{Sample of stars studied in this work.}
\begin{tabular}{llrlcrrl}
\hline
\hline
Star name  & Alternative & Vmag    & Instrument & S/N       & Companion   & Companion               & Companion \\
           & name        &         &            & @5000 \AA\ & period [d]  & mass [M$_\mathrm{Jup}$] & reference \\
\hline
\multicolumn{3}{l}{Hot-BD host stars} \\
\hline
HATS-70	&	TIC 98545929	&	12.57	&	FEROS	&	125	&	1.89	&	12.9$^{+1.8}_{-1.6}$	&	R1	\\
TOI-503	&	BD+13 1880	&	9.43	&	GRACES	&	230	&	3.68	&	53.7$\pm$1.2	&	R2	\\
CoRoT-3	&	TIC 392353449	&	13.29	&	HARPS	&	85	&	4.26	&	21.66$\pm$1.00	&	R3	\\
KELT-1	&	TOI-1476	&	10.70	&	HARPS-N	&	105	&	1.22	&	27.38$\pm$0.93	&	R4	\\
TOI-681	&	TIC 410450228	&	10.89	&	REOSC	&	180	&	15.78	&	88.7$^{+2.5}_{-2.3}$	&	R5	\\
TOI-587	&	HD 74162	&	7.80	&	REOSC	&	330	&	8.04	&	81.1$^{+7.1}_{-7.0}$	&	R5	\\
KELT-25	&	TOI-626		&	9.66	&	FEROS	&	290	&	4.40	&	$<$ 64	&	R6	\\
\hline
\multicolumn{3}{l}{Hot-Jupiter host stars} \\
\hline
WASP-33	&	HD 15082	&	8.30	&	HIRES	&	250	&	1.22	&	$<$ 4.1	&	R7	\\
WASP-167&	KELT-13		&	10.50	&	HARPS	&	205	&	2.02	&	$<$ 8.0	&	R8	\\
WASP-189&	HR 5599		&	6.62	&	HARPS	&	1105	&	2.72	&	1.99$^{+0.16}_{-0.14}$	&	R9	\\
KELT-9	&	HD 195689	&	7.59	&	HARPS-N	&	550	&	1.48	&	2.88$\pm$0.84	&	R10	\\
KELT-17	&	BD+14 1881	&	9.29	&	REOSC	&	210	&	3.08	&	1.31$^{+0.28}_{-0.29}$	&	R11	\\
KELT-20	&	MASCARA-2	&	7.59	&	HARPS-N	&	450	&	3.47	&	$<$ 3.382	&	R12	\\
MASCARA-1&	HD 201585	&	8.30	&	HARPS-N	&	560	&	2.15	&	3.7$\pm$0.9	&	R13	\\
HAT-P-49&	HD 340099	&	10.33	&	SOPHIE	&	135	&	2.69	&	1.730$\pm$0.205	&	R14	\\
MASCARA-4&	HD 85628	&	8.19	&	FEROS	&	180	&	2.82	&	3.1$\pm$0.9	&	R15	\\
HAT-P-70&	HD 287325	&	9.47	&	FEROS	&	210	&	2.74	&	$<$ 6.78	&	R16	\\
KELT-19A&	BD+07 1721	&	9.89	&	GRACES	&	240	&	4.61	&	$<$ 4.1	&	R17	\\
HAT-P-69&	TYC 215-1594-1	&	9.77	&	GRACES	&	220	&	4.79	&	3.58$^{+0.58}_{-0.58}$	&	R16	\\
\hline
\end{tabular}
\tablebib{
Companions data: R1 \citep{zhou19a}, R2 \citep{subjak20}, R3 \citep{deleuil08}, R4 \citep{siverd12}, R5 \citep{grieves21},
R6 \citep{rodriguez20}, R7 \citep{collier10}, R8 \citep{temple17}, R9 \citep{lendl20}, R10 \citep{gaudi17},
R11 \citep{zhou16}, R12 \citep{lund17}, R13 \citep{talens17}, R14 \citep{bieryla14}, R15 \citep{dorval20}, R16 \citep{zhou19b}, R17 \citep{siverd18}.
}
\label{tab.sample}
\end{table*}

We have an ongoing observational campaign aimed to study early-type stars with planets.
The spectra of the stars TOI-503, KELT-19A and HAT-P-69 were acquired through
the Gemini Remote Access to CFHT ESPaDOnS Spectrograph (GRACES).
This device takes advantage of the high-resolution 
ESPaDOnS\footnote{Echelle SpectroPolarimetric Device for the Observation of Stars}
spectrograph, located at the Canada-France-Hawaii Telescope (CFHT) and fed by an
optical fiber connected to the 8.1-m Gemini North telescope at Maunakea, Hawaii.
We used the 1-fiber object-only observing mode, which provides an average
resolving power of $\sim$67500 between 4500 and 
8500 \AA\footnote{http://www.gemini.edu/sciops/instruments/graces/spectroscopy/spectral-range-and-resolution}.
The observations were taken during the observing runs 2021B and 2022A
(Programs ID: GN-2021B-Q-106 and GN-2022A-Q-305, PI: Carlos Saffe).
The final spectral coverage is 4050-9000 \AA, and the signal-to-noise
ratio (S/N) per pixel resulted $\sim$230 measured at $\sim$5000 {\AA} in the combined spectra.
GRACES spectra were reduced using the code OPERA\footnote{Open source Pipeline for ESPaDOnS Reduction
and Analysis} \citep{martioli12}. More recent documentation on OPERA can be found
at the ESPECTRO project webpage\footnote{http://wiki.lna.br/wiki/espectro}.

The spectra of the stars HATS-70, KELT-25, MASCARA-4 and HAT-P-70 
were taken with the Fiber-fed Extended Range Optical Spectrograph (FEROS), which provides a
high-resolution (R$\sim$48000) spectra when illuminated via the 2.0 arcsec aperture
on the sky in the unbinned mode.
FEROS is installed in the Max Planck Gesselschaft (MPG) 2.2-m telescope
at the European Southern Observatory (ESO) in La Silla, Chile.
The data were taken during the observing runs 2021B and 2022A
(Programs ID: 0108.A-9012(A) and 0109.A-9023(A), PI: Marcelo Jaque Arancibia),
including three individual spectra per object, followed by a ThAr lamp 
in order to obtain an appropriate wavelength solution.
The spectra were reduced using the FEROS Data Reduction
System\footnote{https://www.eso.org/sci/facilities/lasilla/instruments/feros/tools/DRS.html} (DRS).
The spectral coverage resulted between 3800-8600 \AA, approximately, and S/N per pixel measured
at $\sim$5000 \AA~resulted in $\sim$210.

In addition, the spectra of the stars TOI-587, TOI-681 and KELT-17 were obtained at the Complejo Astron\'omico
El Leoncito (CASLEO) during the observing runs 2019A, 2021B and 2022A (PI: Carlos Saffe). 
We used the \emph{Jorge Sahade} 2.15-m telescope equipped with a REOSC echelle
spectrograph\footnote{On loan from the Institute d'Astrophysique de Liege, Belgium.} and a TEK 1024$\times$1024 CCD detector.
The REOSC spectrograph uses gratings as cross-dispersers. We used a grating with 400 lines mm$^{-1}$, which provides
a resolving power of $\sim$ 13000 covering the spectral range of $\lambda\lambda$3800--6500.
Three individual spectra for each object were obtained and then combined, reaching a final S/N per pixel
of $\sim$240 measured at $\sim$5000 \AA.
The data were reduced with Image Reduction and Analysis Facility (IRAF)\footnote{IRAF is distributed by
the National Optical Astronomical Observatories, which is operated by the Association of Universities for
Research in Astronomy, Inc., under a cooperative agreement with the National Science Foundation.} 
following the standard recipe for echelle spectra (i.e., bias and flat corrections, 
order-by-order normalization, scattered light correction, etc.).

Finally, we used archive spectra for the case of HARPS, HARPS-N, HIRES and SOPHIE spectrographs.
General characteristics of these instruments are shown in the Table \ref{table.spectrographs},
including the resolving power, CCD detector, pixel size, telescope and approximate wavelength range.
The reduction was performed by using the Data Reduction Software (DRS) pipeline for the case of HARPS and
HARPS-N spectra\footnote{https://www.eso.org/sci/facilities/lasilla/instruments/harps/doc.html},
using the reduction package MAKEE 3 with HIRES spectra\footnote{http://www.astro.caltech.edu/~tb/makee/},
and the DRS pipeline with SOPHIE spectra\footnote{http://www.obs-hp.fr/guide/sophie/sophie-eng.shtml\#drs}.
The continuum normalization and other operations (such as Doppler correction and combining spectra)
were performed using IRAF.

\begin{table}
\centering
\caption{General characteristics of the spectrographs used in this work.}
\scriptsize
\begin{tabular}{lrcclc}
\hline
Instrument & R & CCD        & Pixel  & Telescope & Approx.     \\
           &   & Detector   & size   &           & wave. range \\
\hline
FEROS   &  48000 &  2k x 4k &  15 $\mu$m   & MPG 2.2m      & 3800 - 8600 \\
REOSC   &  13000 &  1k x 1k &  24 $\mu$m   & CASLEO 2.15m  & 3800 - 6500 \\
GRACES  &  67000 &  2k x 4k &  13.5 $\mu$m & Gemini 8.2m   & 4050 - 9000 \\
HARPS   & 115000 &  4k x 4k &  15 $\mu$m   & La Silla 3.6m & 3800 - 6800 \\
HARPS-N & 115000 &  4k x 4k &  15 $\mu$m   & TNG 3.6m      & 3800 - 6800 \\
HIRES   &  67000 &  2k x 4k &  15 $\mu$m   & Keck 10m      & 3750 - 9000 \\
SOPHIE  &  75000 &  4k x 2k &  15 $\mu$m   & OHP 1.93m     & 3900 - 6800 \\
\hline
\end{tabular}
\normalsize
\label{table.spectrographs}
\end{table}

\section{Stellar parameters and abundance analysis}

Stellar parameters were determined similarly to previous works.
The effective temperature T$_{\rm eff}$ and superficial gravity {$\log g$} were first estimated by using the
Str\"omgren uvby$\beta$ mean photometry of \citet{hauck-mermilliod98} or by taking previously published results.
We applied the calibration of \citet{napi93} within the program TempLogG \citep{kaiser06} and
derredenned colors according to \citet{domingo-figueras99}, in order to derive the fundamental parameters.
These initial values were then refined (when necessary and/or possible) by enforcing 
excitation and ionization balances of the iron lines. The same strategy was previously
applied in the literature with early-type stars \citep[e.g., ][]{saffe-levato14,saffe21,alacoria22}.
The values derived in this way are listed in the Table \ref{table.params}, 
including the values taken from our previous work \citep{saffe21}.
The resulting average dispersions are $\sim$186 K and $\sim$0.15 dex for T$_{\rm eff}$ and {$\log g$}, respectively.

\begin{table*}
\centering
\caption{Fundamental parameters adopted for the stars in this work.}
\begin{tabular}{lcccrc}
\hline
Star & T$_{\rm eff}$ & $\log g$ & v$_\mathrm{micro}$ & $v\sin i$     & Reference\\
     &  [K]          &  [dex]   & [km s$^{-1}$]      & [km s$^{-1}$] & \\
\hline
HATS-70	&	7905 $\pm$ 180	&	4.17 $\pm$ 0.12	&	3.28 $\pm$ 0.82	&	39.7 $\pm$ 1.2	&	this work	\\
TOI-503	&	7790 $\pm$ 140	&	4.08 $\pm$ 0.10	&	3.24 $\pm$ 0.81	&	27.2 $\pm$ 0.9	&	this work	\\
CoRoT-3	&	6785 $\pm$ 150	&	4.25 $\pm$ 0.15	&	1.92 $\pm$ 0.48	&	17.8 $\pm$ 0.3	&	this work	\\
KELT-1	&	6720 $\pm$ 210	&	4.23 $\pm$ 0.22	&	1.81 $\pm$ 0.45	&	50.3 $\pm$ 1.7	&	this work	\\
TOI-681	&	7490 $\pm$ 160	&	4.27 $\pm$ 0.14	&	2.99 $\pm$ 0.75	&	43.8 $\pm$ 1.8	&	this work	\\
TOI-587	&	10250 $\pm$ 300	&	4.25 $\pm$ 0.13	&	1.18 $\pm$ 0.30	&	31.5 $\pm$ 0.8	&	this work	\\
KELT-25	&	8100 $\pm$ 190	&	3.86 $\pm$ 0.18	&	3.31 $\pm$ 0.83	&	112.9 $\pm$ 4.2	&	this work	\\
WASP-33	&	7373 $\pm$ 164	&	4.14 $\pm$ 0.20	&	2.86 $\pm$ 0.71	&	82.5 $\pm$ 2.0	&	S21 \& this work	\\
WASP-167&	6770 $\pm$ 210	&	4.05 $\pm$ 0.14	&	1.90 $\pm$ 0.47 &	52.0 $\pm$ 2.0	&	S21 \& this work	\\
WASP-189&	7946 $\pm$ 136	&	3.85 $\pm$ 0.12	&	3.30 $\pm$ 0.82	&	95.2 $\pm$ 3.3	&	S21 \& this work	\\
KELT-9	&	9329 $\pm$ 118	&	4.00 $\pm$ 0.14	&	2.27 $\pm$ 0.57	&	113.5 $\pm$ 4.6	&	S21 \& this work	\\
KELT-17	&	7471 $\pm$ 210	&	4.20 $\pm$ 0.14	&	2.50 $\pm$ 0.63	&	43.0 $\pm$ 0.6	&	S21 \& this work	\\
KELT-20	&	8652 $\pm$ 160	&	4.11 $\pm$ 0.20	&	3.03 $\pm$ 0.76	&	111.0 $\pm$ 3.1	&	S21 \& this work	\\
MASCARA-1&	7687 $\pm$ 238	&	4.09 $\pm$ 0.14	&	3.17 $\pm$ 0.79	&	99.0 $\pm$ 3.9	&	S21 \& this work	\\
HAT-P-49&	6730 $\pm$ 234	&	4.02 $\pm$ 0.14	&	1.82 $\pm$ 0.46	&	14.8 $\pm$ 0.2	&	S21 \& this work	\\
MASCARA-4&	7810 $\pm$ 165	&	3.98 $\pm$ 0.12	&	3.25 $\pm$ 0.81	&	42.3 $\pm$ 1.3	&	this work	\\
HAT-P-70&	8450 $\pm$ 195	&	4.08 $\pm$ 0.14	&	3.19 $\pm$ 0.80	&	103.5 $\pm$ 4.1	&	this work	\\
KELT-19A&	7500 $\pm$ 130	&	4.13 $\pm$ 0.21	&	3.00 $\pm$ 0.75	&	81.5 $\pm$ 2.1	&	this work	\\
HAT-P-69&	7750 $\pm$ 245	&	4.15 $\pm$ 0.09	&	3.21 $\pm$ 0.80	&	76.1 $\pm$ 1.7	&	this work	\\
\hline
\end{tabular}
\tablebib{S21 \citep{saffe21}. }
\normalsize
\label{table.params}
\end{table*}

The projected rotational velocities, $v\sin i,$ were estimated by fitting most \ion{Fe}{I} and
\ion{Fe}{II} lines in the spectra. 
Synthetic spectra were calculated using the program SYNTHE \citep{kurucz-avrett81} together with
ATLAS12 \citep{kurucz93} model atmospheres, convolved with a rotational profile (using the Kurucz’s
command {\it{rotate}}) and with an instrumental profile for each spectrograph (using the command {\it{broaden}}).
Microturbulence velocity v$_\mathrm{micro}$ was estimated as a function of T$_{\rm eff}$ following the
formula of \citet{gebran14}, which is valid for $\sim$6000 K $<$ T$_{\rm eff}$ $<$ $\sim$10000 K.
We adopt for v$_\mathrm{micro}$ an uncertainty of $\sim$25 $\%$, as suggested by \citet{gebran14},
and then this uncertainty was taken into account in the abundance error calculation.

We applied an iterative procedure to determine the chemical abundances for the stars in our sample.
In a first step, we computed an ATLAS12 \citep{kurucz93} model atmosphere adopting initially solar
abundances from \citet{asplund09}. The corresponding abundances are then obtained by fitting
the observed spectra with the program SYNTHE \citep{kurucz-avrett81}.
With the new abundance values, we derived a new model atmosphere and restarted the process again.
In each step, opacities were calculated for an arbitrary composition and v$_\mathrm{micro}$ using the opacity
sampling (OS) method, similar to previous works \citep{saffe20,saffe21,alacoria22}.
If necessary, T$_{\rm eff}$ and $\log g$ were refined to achieve the balance of \ion{Fe}{I} and \ion{Fe}{II} lines.
In this way, parameters and abundances are consistently derived using specific opacities rather than solar-scaled values.
Possible differences originating from the use of solar-scaled opacities instead of an arbitrary
composition were recently estimated for solar-type stars \citep{saffe18,saffe19}.
These differences could became particularly important when modeling chemically peculiar stars,
where solar-scaled models could result in a very different atmospheric structure \citep[e.g. ][]{piskunov-kupka01}
and reach abundance differences up to 0.25 dex \citep{khan-shulyak07}.

\begin{figure}
\centering
\includegraphics[width=8cm]{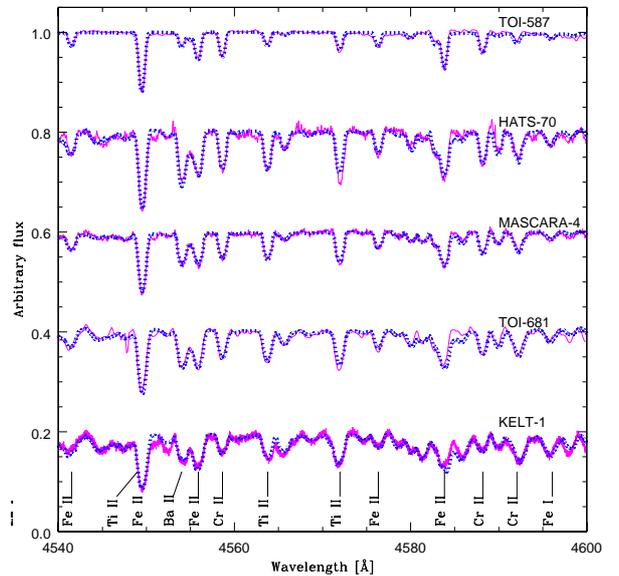}
\caption{Example of observed and synthetic spectra (magenta and blue dotted lines)
for some stars in our sample, sorted by $v\sin i$.}
\label{fig.spectra.example}%
\end{figure}

We derived the chemical abundances for 25 different species, including
\ion{C}{I}, \ion{O}{I}, \ion{Na}{I}, \ion{Mg}{I}, \ion{Mg}{II},
\ion{Al}{I}, \ion{Al}{II}, \ion{Si}{II}, \ion{Ca}{I}, \ion{Ca}{II}, 
\ion{Sc}{II}, \ion{Ti}{II}, \ion{Cr}{II}, \ion{Mn}{I}, 
\ion{Fe}{I}, \ion{Fe}{II}, \ion{Ni}{II}, \ion{Co}{I}, \ion{Zn}{I}, \ion{Sr}{II},
\ion{Y}{II}, \ion{Zr}{II}, \ion{Ba}{II}, \ion{Nd}{II} and \ion{Eu}{II}. 
The laboratory data and atomic line list used in this work are described in \citet{saffe21}.
Figure \ref{fig.spectra.example} presents an example of observed and synthetic spectra
(magenta and blue dotted lines) for some stars in our sample.

The uncertainty in the abundance values was estimated considering different sources.
We estimated the measurement error e$_{1}$ from the line-to-line dispersion
as $\sigma/\sqrt{n}$, where $\sigma$ is the standard deviation and n is the number of lines.
For elements with only one line, we adopted for $\sigma$ the standard deviation of the iron lines.
Then, we determined the contribution to the abundance error due to the uncertainty 
in stellar parameters. We modified T$_{\rm eff}$, $\log g$ and v$_\mathrm{micro}$ by their
uncertainties and recalculated the abundances, obtaining the corresponding
differences e$_{2}$, e$_{3}$ and e$_{4}$\footnote{We adopt a minimum of 0.01 dex for these errors.}.
Finally, the total error e$_{tot}$ was estimated as the quadratic sum of e$_{1}$, e$_{2}$, e$_{3}$ and e$_{4}$.
The abundances with their total error e$_{tot}$, the individual errors e$_{1}$ to e$_{4}$ and the
number of lines n, are presented in the Tables \ref{tab.abunds.CoRoT-3} to \ref{tab.abunds.HAT-P-69}
of the Appendix.


In general, departures from local thermodynamic equilibrium (LTE) are pronounced in stars
with high temperature and with low gravity and metallicity.
Then, some particular transitions should be taken with caution for early-type stars.
For the case of \ion{O}{i}, \citet{sitnova13} estimated non-LTE (NLTE) corrections for stars with T$_{\rm eff}$
between 5000 K and 10000 K. They showed that NLTE effects lead to a strengthening of \ion{O}{i} lines.
Then, we estimated \ion{O}{i} NLTE abundance corrections for the IR triplet at 7771 \AA\ and for the line 5328.98 \AA\
for the stars in our sample, depending on their fundamental parameters.
If available, we prefer to use the line at 5328.98 \AA\ which presents considerably lower NLTE effects than the IR triplet
\citep{sitnova13}.

NLTE effects could also be present in species such as \ion{Mg}{i} and \ion{Mg}{ii} \citep[e.g. ][]{przybilla01,bergemann17,alexeeva18}.
\citet{przybilla01} found that the intense line \ion{Mg}{II} 4481 \AA\ systematically
yields notably higher abundances due to NLTE effects (between 0.2 - 0.8 dex for early-type stars),
while \citet{alexeeva18} estimate that the average difference between \ion{Mg}{I} and \ion{Mg}{II}
diminish from $\sim$0.23 dex in LTE to $\sim$0.09 dex in NLTE. This could explain, at least in part, the
higher abundances of \ion{Mg}{II} compared to \ion{Mg}{I} observed in some stars.
However, the authors also caution that the difference \ion{Mg}{II} $-$ \ion{Mg}{I} even in NLTE
could amount up to $\sim$0.24 dex for metal-poor stars, for a reason that requires future investigation.
We estimated \ion{Mg}{I} NLTE abundance corrections following \citet{bergemann17},
who calculated Mg abundance departures between LTE and NLTE by using 1D and spatially averaged $<$3D$>$ model atmospheres.
They showed that \ion{Mg}{I} NLTE effects grow with increasing T$_{\rm eff}$, and decreasing {$\log g$} and [Fe/H],
that is to say, they are sensitive to stellar parameters.
The lower NLTE effects of \ion{Mg}{I} lines compared to the intense line \ion{Mg}{II} 4481 \AA, 
result in more reliable abundance values for Mg.
In addition to these considerations, we prefer to avoid the line \ion{Mg}{I} 5167.32 \AA\ which is often
strongly blended with \ion{Fe}{I} 5167.49 \AA.

For the case of the calcium lines, we estimated NLTE abundance corrections following \citet{mashonkina07},
who determined abundances in LTE and NLTE.
They calculated a detailed Ca model atom and then obtained statistical equilibrium populations
by adopting a plane-parallel homogeneous MAFAGS model atmosphere.
We also note that, beyond NLTE effects, we prefer to avoid the \ion{Ca}{II} line 3968.47 \AA\
which is strongly blended with the Balmer line H$\epsilon$ in many early-type stars.

\section{Discussion}

\subsection{Defining Am stars using abundances}

As explained in the Introduction, a number of authors consider Am stars as objects showing 
underabundances of Ca and/or Sc, together with overabundances of Fe-peak and heavier elements
\citep{preston74,fossati07,fossati09,catanzaro19}.
In this work we consider "Am stars" those early-type stars showing the mentioned characteristics.
However, following the work of \citet{conti70}, other authors prefer a wider definition
for the class, showing underabundances of Ca and/or Sc, and/or overabundances of the Fe group
and heavier elements \citep{conti70,lane-lester87,adelman94,adelman97,yuce-adelman14}. 
As a consequence of the wider definition, it would be possible to increase the number of Am stars found.
Those objects satisfying the wider definition rather than the first definition, are referred in this
work to as "possible Am" stars.
We will estimate the incidence of Am stars using both definitions, "Am stars" and "possible Am stars",
where the reader could decide the more appropriate for their own preference.

\subsection{Am stars in our sample}

\begin{figure}
\centering
\includegraphics[width=8cm]{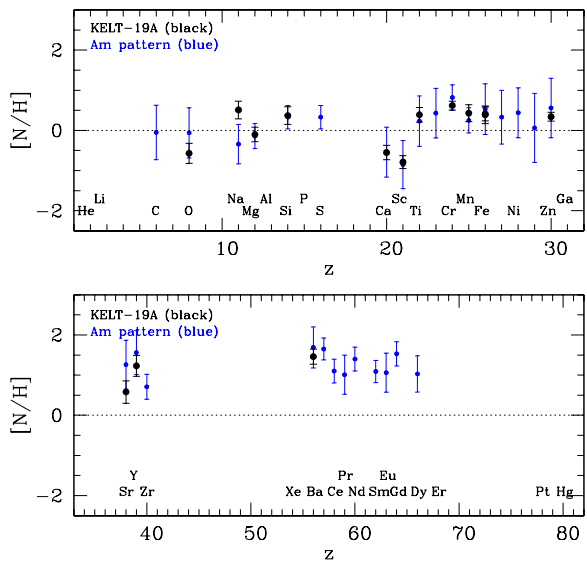}
\caption{Abundances of the star KELT-19A (black), compared to the average pattern
of Am stars (blue).
The panels correspond to elements with z$<$32 and z$>$32.}
\label{pattern.KELT-19A}%
\end{figure}

We discuss in this Section those stars in our sample that could be related to Am stars
in general, that is to say, Am stars and possible Am stars.
We compared the abundances of the stars in our sample with an average pattern of 62 Am stars derived by \citet{catanzaro19}.
In particular, we present in Fig. \ref{pattern.KELT-19A} the abundances of the star KELT-19A
(black) compared with those of Am stars (blue).
The square bracket notation [N/H] corresponds to abundance values relative to the Sun\footnote{[N/H] $=$ log(N/H)$_{star}$ $-$ log(N/H)$_{Sun}$},
with solar values taken from \citet{asplund09}.
The Figure includes two panels, corresponding to elements with atomic number z$<$32 and z$>$32.
Ca and Sc show subsolar values (with Ca slightly higher than Sc),
while Fe and other metals show overabundances that resemble those of Am stars.
An apparent disagreement is Na (z$=$11) showing suprasolar rather than subsolar values,
however some Am stars also present suprasolar values
\citep[e.g. HD 267 and HD 176716, ][]{catanzaro19}.
Then, abundance values agree with the definition of Am stars given
in the previous Section.
\citet{siverd18} found a transiting hot-Jupiter planet orbiting around this star,
and suggest its possible Am nature by noticing in the spectra an enhancement in the
metallic lines together with a calcium deficiency.
Then, we confirm the previous suggestion of \citet{siverd18} and
consider that KELT-19A could be safely identified as an Am star.

\begin{figure*}
\centering
\includegraphics[width=8cm]{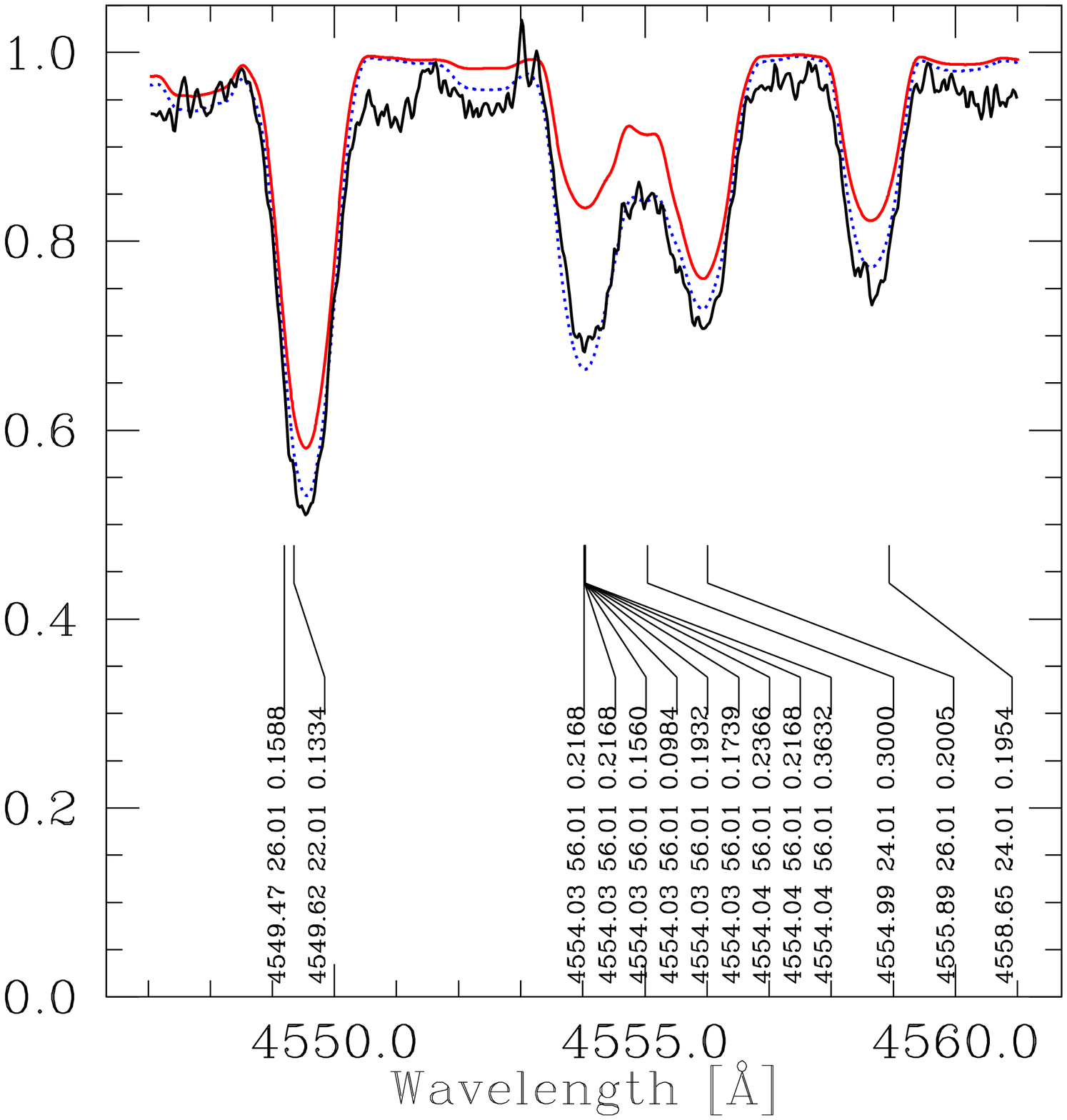}
\includegraphics[width=8cm]{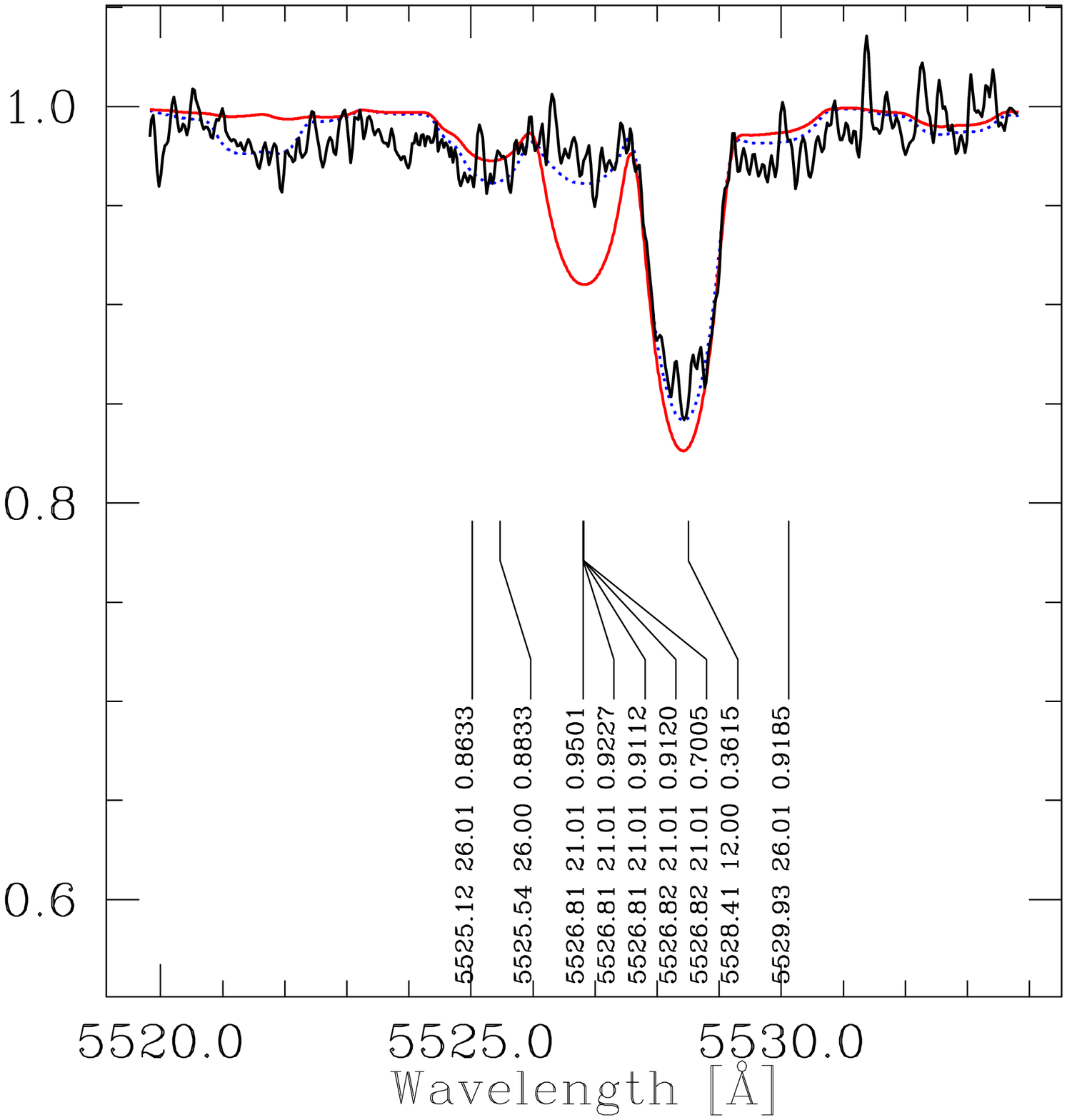}
\caption{Observed spectra of the star HATS-70 (black), compared to synthetic spectra with our abundances
(blue dotted) and adopting a solar composition (red).
The two panels correspond to two different spectral regions.
Line identifications present the wavelength, atomic number (with ionization state as decimals), and intensity (between 1 and 0).
}
\label{spectra1.HATS-70}%
\end{figure*}

\citet{zhou19a} first identified a transiting companion to the early-type star HATS-70,
with a mass of 12.9 M$_\mathrm{Jup}$ and a period of 1.89 d. They performed an initial estimation
of the stellar parameters by fitting the observed spectra with a grid of synthetic spectra with steps
of 0.5 dex in [Fe/H]. Then, the final stellar parameters adopted were estimated
within a global analysis via a spectral energy distribution (SED) fit.
However, the authors caution that the metallicity is poorly constrained in the SED fitting
and adopt an approximate near-solar value of [Fe/H]$\sim$0.03 dex (see their Appendix 1).
Hosting a hot-companion, we take the opportunity and performed a detailed
abundance analysis for this early-type star. We present in the Fig. \ref{spectra1.HATS-70}
the observed spectra of the star HATS-70 (black), compared to synthetic spectra with our abundances
(blue dotted) and adopting a solar composition (red). The line identifications are shown using three numbers:
wavelength, atomic number (and ionization state as decimals), and intensity (between 1 and 0).
For instance, the label "3933.68 20.01 0.1234" would correspond to the line \ion{Ca}{ii} 3933.68 \AA.
The two panels show an example of different spectral regions.
The left panel shows that some metals are clearly enhanced, while the right panel shows that
the line \ion{Sc}{ii} 5226 \AA\ is notably weaker than expected for a solar composition.
Then, Fig. \ref{spectra1.HATS-70} suggests that it would be difficult to fit the spectra of HATS-70 adopting a solar composition.

\begin{figure}
\centering
\includegraphics[width=8cm]{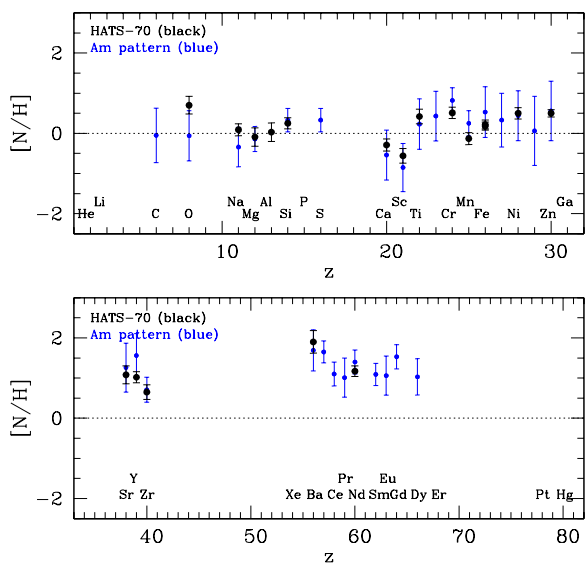}
\caption{Abundances of the star HATS-70 (black), compared to the average pattern
of Am stars (blue).
The panels correspond to elements with z$<$32 and z$>$32.}
\label{pattern.HATS-70}%
\end{figure}

We present in Fig. \ref{pattern.HATS-70} the abundances of the star HATS-70
(black) compared to the average pattern of Am stars (blue).
Most abundance values reasonably agree with those of Am stars.
Ca and Sc present subsolar values (being Ca slightly higher than Sc),
similar to other Am stars.
Fe presents slightly lower abundances than average Am stars, being still suprasolar
([\ion{Fe}{i}/H]$=$0.18$\pm$0.10 dex, [\ion{Fe}{ii}/H]$=$0.22$\pm$0.11 dex).
For example, the Am stars HD 134214, HD 139939 and HD 159545
present a very similar Fe enhancement \citep{catanzaro19}.
The \ion{Ni}{ii} abundance (0.50$\pm$ 0.14 dex) was derived using the line 4015.47 \AA, 
showing an enhancement in good agreement with other Am stars.
The heavy elements (Sr, Y, Zr) and the most heavy elements (Ba, Nd)
show strong overabundances between $\sim$0.5-2.0 dex, corresponding to
a chemically peculiar (non-solar composition) star.
Having subsolar Ca and Sc, together with suprasolar values of most Fe-peak and heavier
elements, HATS-70 fits the previous definition of Am stars.

\begin{figure}
\centering
\includegraphics[width=8cm]{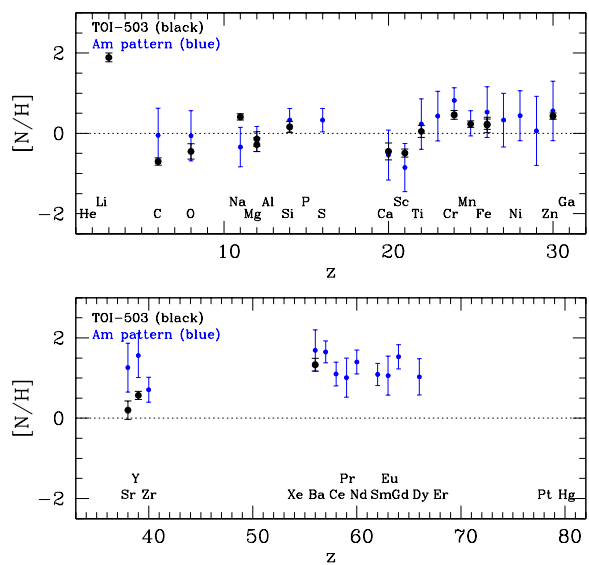}
\caption{Abundances of the star TOI-503 (black), compared to the average pattern
of Am stars (blue).
The panels correspond to elements with z$<$32 and z$>$32.}
\label{pattern.TOI-503}%
\end{figure}

\begin{figure}
\centering
\includegraphics[width=8cm]{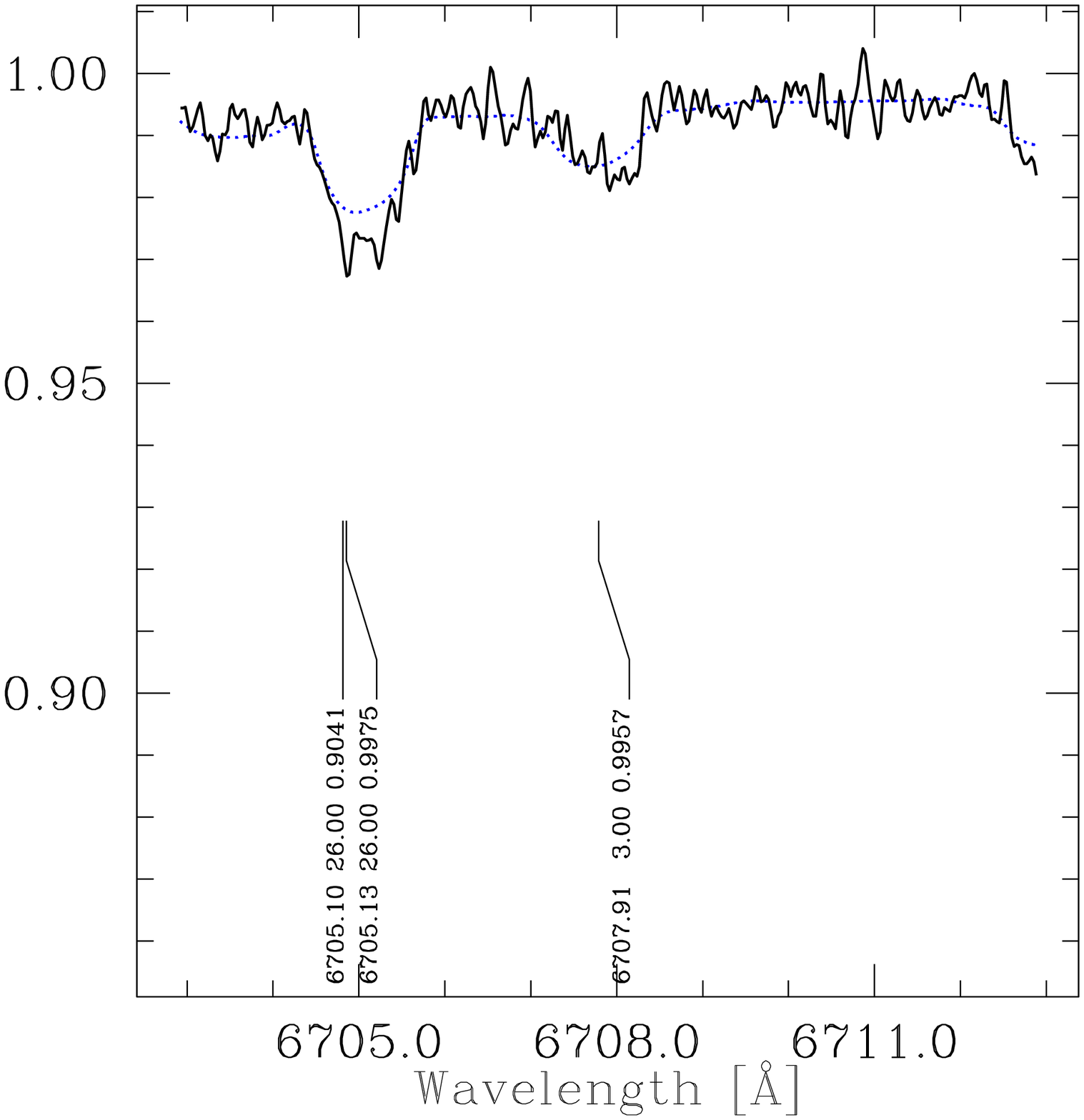}
\caption{Observed and synthetic spectra of the star TOI-503 (black and blue dotted lines) near the \ion{Li}{i} line 6707.8 \AA.
Line identifications are similar to those of Fig. \ref{spectra1.HATS-70}.  }
\label{spectra1.TOI-503.Li6707}%
\end{figure}

S20 reported the discovery of TOI-503b, the first BD known to orbit a metallic-lined A star (Am star).
They derived abundances for some key elements of the host star (Fe, Ni, Ca, Sc and Mg) which suggest its Am nature.
We included this star in our sample and performed a detailed abundance analysis including as many species as possible.
We present in Fig. \ref{pattern.TOI-503} the abundances of the star TOI-503
(black) compared to the average pattern of Am stars (blue).
The general behavior of the chemical species strongly resembles those of Am stars,
confirming the previous suggestion of S20 about the Am nature of this object.
We note relatively low abundances for C and O compared to average Am stars. 
We corrected O abundances (derived from the oxygen triplet near $\sim$7774 \AA) of NLTE effects
with an average correction of $\sim$-0.4 dex.
Also, we note that the \ion{Li}{i} line 6707.8 \AA\ is relatively weak although still present in the spectra of TOI-503.
We present in the Fig. \ref{spectra1.TOI-503.Li6707} observed and synthetic spectra of the star TOI-503 (black and blue dotted lines) in a spectral region near the Li line.
The corresponding abundance derived for lithium ([\ion{Li}{i}/H]$=$1.98$\pm$0.11 dex)
resulted similar to other Am stars \citep[e.g. ][]{iliev98,stateva12b}.
Then, showing underabundances of both Ca and Sc, together with suprasolar values of
most Fe-peak and heavier elements, we consider TOI-503 as Am star.

We included in our sample of early-type stars with hot-companions the
star KELT-17. The Am nature of this object was reported in our previous works \citep[][]{saffe20,saffe21}
using the same model atmospheres 
and laboratory data, showing subsolar Ca and Sc, together with suprasolar Fe-peak elements.
Then, in the complete sample of 19 early-type stars with hot-companions studied in this work,
we detected 4 Am stars: KELT-19A, HATS-70, TOI-503 and KELT-17.

\begin{figure}
\centering
\includegraphics[width=8cm]{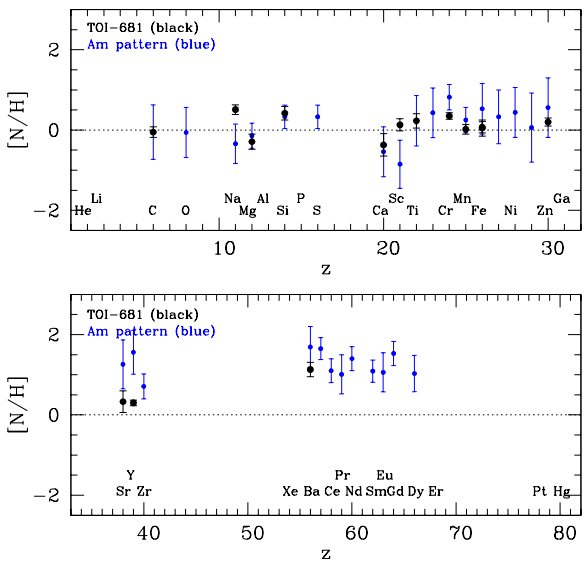}
\caption{Abundances of the star TOI-681 (black)
compared to the average pattern of Am stars (blue).
Upper and lower panels correspond to elements with z$<$32 and z$>$32.}
\label{pattern.TOI-681}%
\end{figure}

\begin{figure}
\centering
\includegraphics[width=8cm]{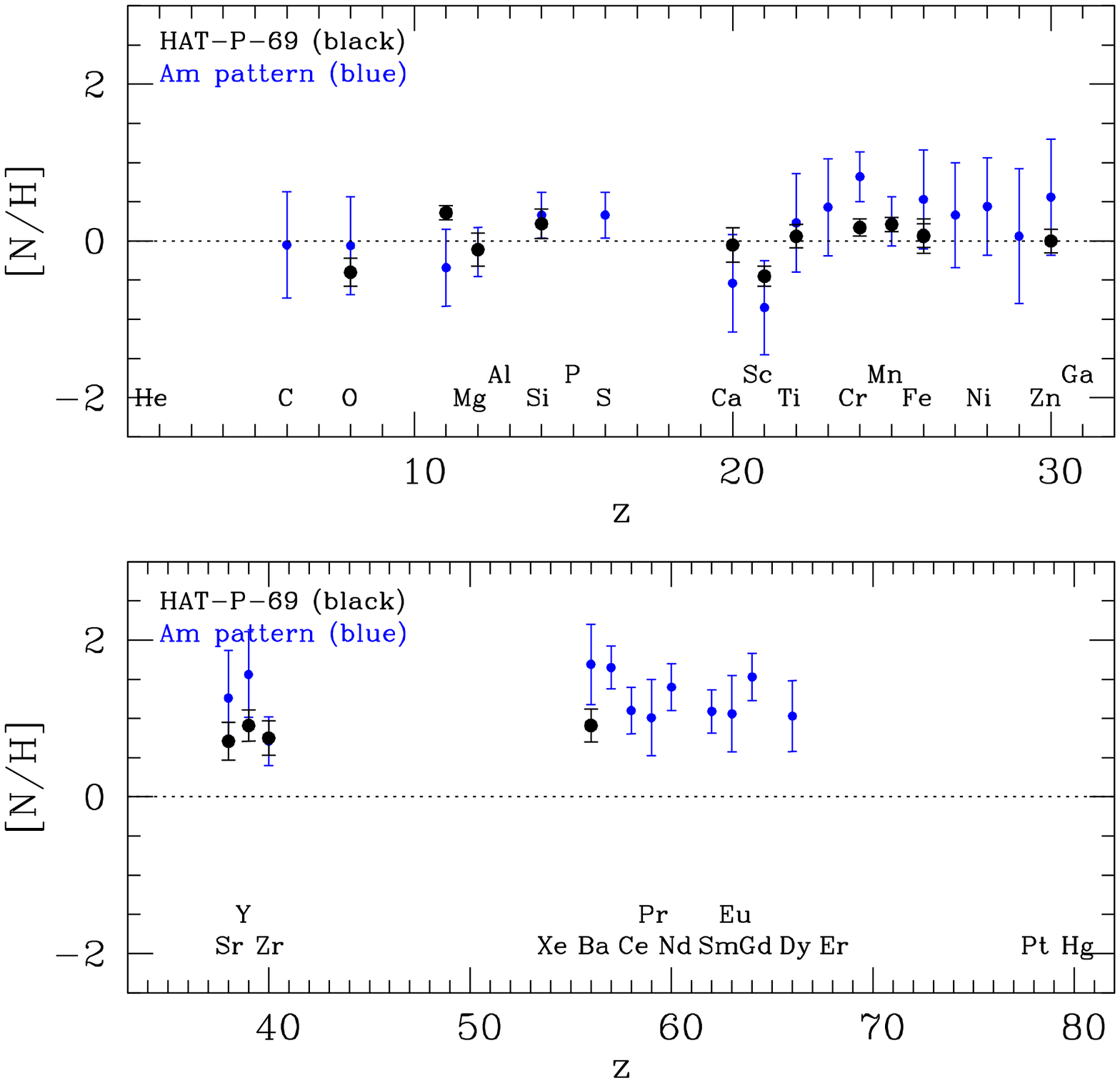}
\caption{Abundances of the star HAT-P-69 (black)
compared to the average pattern of Am stars (blue).
Upper and lower panels correspond to elements with z$<$32 and z$>$32.}
\label{pattern.HAT-P-69}%
\end{figure}

\begin{figure}
\centering
\includegraphics[width=8cm]{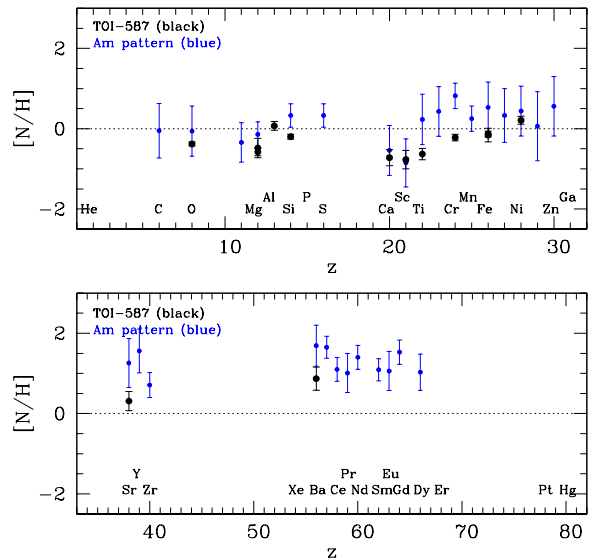}
\caption{Abundances of the star TOI-587 (black)
compared to the average pattern of Am stars (blue).
Upper and lower panels correspond to elements with z$<$32 and z$>$32.}
\label{pattern.TOI-587}%
\end{figure}

Next, we discuss some stars with a less clear Am identification.
We present in the Figs. \ref{pattern.TOI-681}, \ref{pattern.HAT-P-69} and \ref{pattern.TOI-587}.
the chemical pattern of the stars TOI-681, HAT-P-69 and TOI-587 compared to
the average pattern of Am stars.
The star TOI-681 presents subsolar Ca, solar Fe, together with enhanced Ti, Cr and some
rare earths. Then, following the definition given in the previous section,
we consider this object as a possible Am star.
As we can see in the Fig. \ref{pattern.HAT-P-69}, the star HAT-P-69 presents subsolar Sc,
solar Fe-peak species and some enhancements in heavier elements (Sr, Y, Zr, Ba).
These characteristics indicate that HAT-P-69 could be identified as a possible Am star.
The star TOI-587 presents subsolar Ca and Sc, however most elements also present subsolar
abundance values (O, Mg, Si, Ti, Cr and Fe), together with some rare-earths enhancements.
Its chemical pattern shows mostly a metal-poor star, with the exception of few elements.
So we prefer to be cautious and not relate this object to the Am class.
Finally, in the complete sample of 19 stars, we have identified 4 Am stars and, 
adopting a wider definition for the class, 2 possible Am stars (TOI-681 and HAT-P-69).

\subsection{Other chemical patterns in our sample}

\begin{figure}
\centering
\includegraphics[width=8cm]{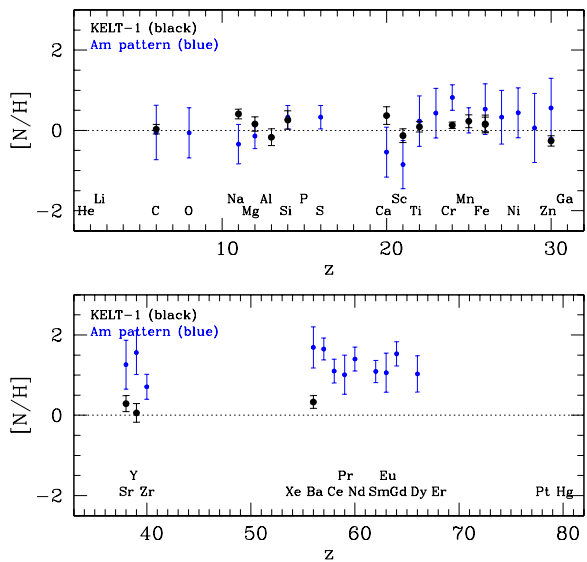}
\caption{Abundances of the star KELT-1 (black)
compared to the average pattern of Am stars (blue).
Upper and lower panels correspond to elements with z$<$32 and z$>$32.}
\label{pattern.KELT-1}%
\end{figure}

We discuss in this section the chemical pattern of stars that 
do not seem to be Am stars.
We present in the Fig. \ref{pattern.KELT-1} the chemical pattern
of the star KELT-1. This object
presents slight overabundances in Si, Cr and Mn; Fe resulted almost solar within errors;
Ca is overabundant and Sc slightly deficient (although solar within the errors).
Then, although KELT-1 presents slight suprasolar values in some species,
the general behavior of the abundances shown in Fig. \ref{pattern.KELT-1}
does not seem to be related to Am stars.

Within the group of stars with abundances determined in this work,
some of them resulted with solar or near-solar composition
(CoRoT-3 and KELT-25),
as we can see in the Tables \ref{tab.abunds.CoRoT-3} to \ref{tab.abunds.HAT-P-69}
of the Appendix.
In the sample we also found stars with slightly subsolar abundances
(HAT-P-70 and MASCARA-4).
Then, we should add to this group 7 stars with chemical patterns
derived in our previous work \citep{saffe21} that do not seem to display an Am spectra 
(WASP-33, WASP-167, WASP-189, KELT-9, KELT-20, MASCARA-1 and HAT-P-49). 
In this way, in the complete sample of 19 early-type stars
hosting hot-companions, we identified 4 Am stars and 2 possible Am stars, while the 13
remaining objects do not seem to be related to this class.

\subsection{Tidal effects between Am stars and low-mass companions?}

We explore in this section the possible tidal effects between the Am stars detected and their hot-companions.
S20 discovered TOI-503, the first Am star which hosts a very close (a$\sim$0.06 au) BD companion. 
They found that the rotation period of the star is similar to the orbital period of the
BD (P$_{orb}=$3.67 days). The authors tried to study the tidal evolution and circularization of the system.
However, as discussed by the authors (their section 4.3), the general uncertainty of tidal evolution models 
difficult to estimate the interaction between the star and their hot companion.
They showed that the tidal circularization and tidal synchronization timescales are notably shorter
for lower values of the tidal quality factor Q', using the tidal model of \citet{jackson08}.
Alternatively, using the tidal model of \citet{zahn77}, they predict timescales one order of magnitude larger
than those using the model of \citet{jackson08}. 
Then, they cannot draw firm conclusions on the tidal evolution of the system (that is to say,
if the star TOI-503 circularized or not the BD’s orbit), given the uncertainty in the appropriate
tidal model and the values of Q'.
However, they do not discard the possibility that the presence of BDs and Am stars could be linked.

\citet{zhou19a} discovered a BD with a mass of 12.9 M$_\mathrm{Jup}$ orbiting the star HATS-70.
They estimate a rotational velocity of the star 5$\sigma$ different
than expected from a spin synchronized with the orbit of the BD companion.
Following the procedure of \citet{matsumura10}, they found that the angular momentum
of HATS-70 does not reach the value required for the Darwin stability criteria.
They also estimated the timescale required by the planet to modify the
stellar spin following \citet{hansen12}, and found a excessively large value.
Then, although the obliquity of the BD is relatively low ($\lambda=$ 8.9$^{+5.6}_{-4.5}$ deg),
current models do not seem to support a strong star-planet synchronization.

\citet{siverd18} found a hot-Jupiter planet orbiting the Am star KELT-19A.
The star belongs to a binary system together with KELT-19B, a late-type star
separated by $\sim$160 au. Although KELT-19A is a slow rotating star, the
authors suggest that its stellar companion is too distant (and its planetary companion likely too low in mass)
to cause a significant tidal braking (although without a numerical estimation of these effects).
Instead, they consider that the slow rotation of the star is probably either primordial or was induced by
a more efficient tidal braking mechanism.
They also modeled the orbital evolution of the planet due to tidal friction using the program POET \citep{penev14},
under the assumption of a non-rotating star, ignoring a larger-distance Type II migration, scattering-induced
migration and other possible hot-Jupiter planets. It only considers a close-in migration due to tidal friction.
However, even under these simplifying assumptions, 
the authors caution that the estimated evolution of the semimajor-axis with time strongly
depends of the stellar tidal quality factor Q', which presents large uncertainties and is difficult
to estimate \citep[e.g. ][]{ogilvie14,petrucci20}.

\citet{stateva12b} searched for possible tidally driven abundance anomalies
in a sample of Am stars than belong to binary systems.
They found that the "metallicism" defined as [Ca/Fe] $=$ [Ca/H] $-$ [Fe/H] \citep[see e.g. ][]{conti70}
seem to slightly increase with increasing orbital eccentricity, while no trend was detected
with orbital period. 
They suggest that a mechanism should stabilize the atmosphere and reduce the mixing,
due to the presence of a stellar companion.
In our case, the small number statistics (only 2 Am stars hosting hot-BDs
and 2 Am stars hosting hot-Jupiter planets) preclude us to reach robust
conclusions.
These correlations would indicate that tidal effects could play a role in the
development of Am stars. 

Current estimations of the tidal evolution (including a number of uncertainties) do not seem to 
support a strong star-planet synchronization between Am stars and their hot-companions.
However, as a consequence of the small separations, a tidal interaction is not ruled out \citep[e.g.][]{mathis13,ogilvie14}.
A close star-planet interaction could also involve additional mechanisms, such as the recently proposed
resonant oscillations in late-type stars \citep{lanza22a,lanza22b}.

\subsection{Searching stellar companions to Am stars}

We explore possible stellar companions to the Am stars found in our sample.
We searched our targets stars in different binary catalogs from the literature \citep{sb9,shaya-olling11,tokovinin-lepine12,andrews17}.
\citet{shaya-olling11} developed a Bayesian method to discover non-random pairs using Hipparcos proper motions and parallaxes,
while \citet{tokovinin-lepine12} constructed a catalog of common-proper motion pairs within 67 pc of the Sun based on the SUPERBLINK
proper-motion survey. \citet{andrews17} performed a search of binaries through a Bayesian formulation
in the Tycho-Gaia DR2 catalogs and derived likelihoods of Keplerian orbits.
We find no record for the Am stars KELT-19A, HATS-70, TOI-503 and KELT-17 in the mentioned binary catalogs.
We also searched the Am stars in the SB9 catalog of spectroscopic binary orbits\footnote{https://sb9.astro.ulb.ac.be/}
\citep{sb9} and found no entry.

In addition, we searched for signs of additional substellar and/or stellar companions around the four Am stars
KELT-19A, KELT-17, HATS-70, and TOI-503, by using the 2-min cadence photometric data provided by the TESS mission
\citep{ricker15}.
These publicly available data were analyzed with the tools in the Lightkurve Python package \citep{light18}.
Briefly, we first eliminated the data-points of the transits produced by the planet around each star, then performed
a detrending to remove any remaining systematics in the light curve and, finally, ran the Transit Least Squares code
\citep[TLS, ][]{hippke-heller19},
designed to detect periodic transits from time-series photometry\footnote{Although TLS is optimized to find planetary transits,
it can be used to search for signals of eclipses produced by stellar and substellar companions.}. 
We did not detect any feature that could be clearly attributed to the transit of a new planet or the eclipse 
of a companion around KELT-19A, KELT-17, HATS-70, and TOI-503 .

\citet{siverd18} performed adaptive optics observations of the Am star KELT-19A with the Palomar Observatory Hale Telescope
and found a late-type G9V/K1V companion. The projected separation between both stars is 0".64$\pm$0".03,
which corresponds to $\sim$160 au at a distance of 255$\pm$15 pc. They suggest that the companion is bound
based on radial velocity observations. However, given the separation between both stars, they
consider unlikely that the companion caused the tidal braking of the Am star.
\citet{zhou19a} found no additional stellar companions to the Am star HATS-70 within 2" in the Astralux lucky imaging
observations, nor within 5" in the Gaia DR2 catalog. They also note an agreement between the observed
spectral energy distribution (SED) and single star spectral models, ruling out a possible nearby star.
In addition, searching for fainter stellar blends, they injected and attempt to recover an additional stellar signal in the
out-of-transit line profiles. No companions were found in this case.

S20 performed Y-band imaging observations of the Am star TOI-503 with the Sinistro camera on the Las Cumbres Observatory and
the Siding Spring Observatory (SSO). They confirmed that there are no nearby or background eclipsing binaries
within 2'.5 of the star. In addition, no nearby sources were identified in Gemini/NIRI and Keck/NIRC2 images
taken in the Br-$\gamma$ filter within 2" of the target.
\citet{zhou16} found a transiting planet orbiting the star KELT-17, detected using radial velocities and Doppler tomography.
They discard possible blend scenarios (such as background eclipsing binaries) by the detection of the Doppler tomographic signal.
The authors also performed a spectroscopic follow-up taking multi-epoch spectra of the star
using the WiFeS integral field spectrograph, which help to eliminate the possibility of nearby eclipsing binaries.

\subsection{Are Am stars and hot-low mass companions related?}

As discussed in the previous sections, early-type stars with hot companions show
different chemical patterns, some of them being Am stars.
To determine if the presence of hot-companions could facilitate (in some way) the rise of Am stars, 
it would be interesting to compare with the frequency of Am stars in general.
This comparison should be taken with caution, given the (still) relatively low number of early-type stars with hot-companions detected.  
\citet{smith71} found an incidence of $\sim$31\% of Am stars in a random sample of 70 bright A-type stars.
Then, \citet{abt81} studied a sample of 865 components of visual multiples and found a frequency of $\sim$32\%
in the range A4-F1 (IV and V), and decreasing to $\sim$5\% in the hotter range A0-A3 (IV and V).
More recently, \citet{gray16} studied the spectra of 80447 low-resolution spectra of the LAMOST-Kepler survey
and applied the automatic classification code MKCLASS. In this way, they found an incidence of 36.4\% 
in the spectral range A4-F1.
Then, \citet{qin19} searched for metallic-lined Am stars in more than 200000 low-resolution spectra of the LAMOST DR5 survey.
They used a combination of machine learning algorithms and then manual identifications, and found
a total of 9372 Am stars and 1131 Ap stars\footnote{Ap stars are chemically peculiar objects showing, for instance,
intensified lines of Cr, Sr and Eu \citep[see e.g.][]{ghazaryan18,saffe-levato04}.}.
They found that most Am stars belong to temperature interval between $\sim$7250-8250 K ($\sim$F0-A4)
with an incidence of $\sim$17\%, while in hotter and colder intervals the frequency is lower.
The authors suggest that the lower frequency compared to previous studies \citep[][]{smith71,abt81,gray16}
is due to the strict screening applied by \citet{qin19}.

Chemical peculiarities similar to those of Am stars develop for stars hotter than the Kraft break \citep{kraft67}
near $\sim$6200 K. Beyond this limit, hotter stars present gradually thinner convective envelopes that allow the 
diffusion processes to operate.
The coldest Am star found in the catalog of \citet{ghazaryan18} presents T$_{\rm eff}\sim$6780 K.
These values indicate that stars hotter than $\sim$6750 K, could develop an Am peculiarity.
Our sample includes 4 stars with temperatures similar or lower than this value
(CoRoT-3, KELT-1, WASP-167 and HAT-P-49, see Table \ref{table.params}).
These stars were analyzed for completeness and will be excluded from the incidence calculations.
On the other hand, we also prefer to exclude the hot star TOI-587 from the incidence calculations,
having an estimated T$_{\rm eff}$ of 10250 $\pm$ 300 K
(the hotter Am star in the Ghazaryan's catalog presents T$_{\rm eff} \sim$ 10350 K).

The group of hot-BD hosts studied in this work includes 7 objects, with 4 of them in the approximate
T$_{\rm eff}$ range of Am stars.
Within this group, we found the Am stars HATS-70 and TOI-503, being roughly a fraction of 2/4 ($\sim$50\%).
If we also include the star TOI-681, classified as a possible Am star, the fraction of Am stars between hot-BDs would rise to 3/4 ($\sim$75\%).
In other words, in the group of hot-BD hosts, the fraction of Am stars would be roughly between 50-75\%.
Similarly, the group of hot-Jupiters host stars studied in this work includes 12 objects, where 10 of them present T$_{\rm eff}$ hotter than $\sim$6750 K.
Within this group, we found the Am stars KELT-17 and KELT-19A, being roughly a fraction of 2/10 ($\sim$20\%).
Recent literature works showed that the hot-Jupiter host stars MASCARA-5 \citep{addison21} and KELT-26 \citep{rodriguez20} are also likely Am stars
(they are not included in our sample).
If we include these two likely Am stars from the literature, and also the possible Am star HAT-P-69
identified in this work, the fraction of Am stars would rise to 5/12 ($\sim$42\%).
Then, in the group of hot-Jupiters host stars, the fraction of Am stars would be roughly 20-42\%.
 
If we combine the groups of hot-BD and hot-Jupiters hosts (all the stars in our sample) we would have 19 stars, 
with 14 of them in the approximate T$_{\rm eff}$ range of Am stars. In this group, we detected a total of 4 Am stars,
being roughly a fraction of 4/14 ($\sim$28\%).
If we also include the likely Am stars MASCARA-5, KELT-26 and the possible Am star HAT-P-69,
the fraction would rise to 7/16 ($\sim$44\%).
Then, in the group of hot-low mass companions (including both hot-BDs and hot-Jupiters),
the incidence of Am stars would be roughly 28-44\%.

The number of early-type stars having hot-companions detected is still low, and then,
we stress that the comparisons should be taken with caution.
On the one hand, the incidence of Am stars between hot-BDs is roughly 50-75\%,
which would be higher than the frequency of Am stars in general (31-36\%).
On the other hand, the incidence of Am stars with hot-Jupiters hosts is roughly between 20-42\%,
which is not significantly higher than those of Am stars in general.
If confirmed, these incidence estimations would support the idea that the presence
of hot-BDs could help in the development of Am stars, different to the case of hot-Jupiter planets.
Tidal synchronization calculations indicate an apparent lack of a strong star-planet (or star-BD) effect,
however they still present a number of uncertainties, as discussed in the previous section.
The initial trends found here would imply that the presence of hot-BDs could possibly help to mitigate
the "single Am" problem, different to the case of hot-Jupiter planets.
Then, for the case of Am stars hosting hot-Jupiter planets, other mechanisms should be invoked
to explain the presence of these single Am stars or Am stars in wide binary systems.

If confirmed, the higher incidence of Am stars between stars hosting hot-BDs (50-75\%)
than in stars hosting hot-Jupiter planets (20-42\%), would support the idea that the presence
of hot-BDs could affect their host stars.
This could be related to the higher mass of BDs compared to giant planets, or to a different formation mechanism.
These interesting results should be taken with caution, and would deserve
a further confirmation using a larger sample of stars.

\subsection{On the origin of single Am stars}

A search for eclipsing binaries among Am stars suggests that around 60–70 per cent
of Am stars are spectroscopic binaries \citep{smalley14}, which is consistent
with radial velocity studies \citep{abt-levy85,cp07}.
However, the presence of single Am stars and the presence of Am stars in wide binary systems
cannot be ruled out.
Some works mention the possible existence of single Am stars \citep[e.g. ][]{conti69,rao-abhyankar91}, 
including stars such as HD 29499 \citep{debernardi00}, HD 8801 \citep{henry-fekel05} and KIC 3429637 \citep{murphy12}.
Also, different studies of binarity indicate that there are some apparently single
Am stars that were presumably born with slow rotation \citep{smalley14,balona15a}.
In Sect. 4.5 we showed that the Am stars detected here belong to wide binary systems (KELT-19A)
or seem to be single (KELT-17, HATS-70 and TOI-503).
In the previous Section, we showed that the presence of hot-BDs could possibly play a role
in the development of single Am stars, however this is different for the case of hot-Jupiter planets.
Then, other mechanisms should be explored to explain the origin of these Am stars.

The presence of a magnetic field during pre-main-sequence
is considered an efficient braking mechanism.
However, Am stars only present weak or ultra-weak magnetic fields \citep[e.g., ][]{folsom13,blazere16,blazere20}.
\citet{abt-levy85} suggest that evolutionary effects (such as expansion) and rotational spin-down
is the reason for the Am characteristics in long-period binaries and some single stars.
However, \citet{henry-fekel05} noted that the main-sequence nature of HD 8801 argues against this evolutionary scenario.
Instead, they speculate that some additional factor could be responsible for the slow rotation
of the single Am stars, such as the presence of planetary systems.
\citet{rao-abhyankar91} suggest that rapidly rotating A stars
could develop shell structures at some time, then the star
has to slow down in its shell and exhibit Am characteristics.
However, to our knowledge, this possibility was not further explored.
More recently, \citet{murphy13} suggested the possibility that some slow rotators
that emerge from pre-main sequence without the presence of magnetic fields,
with the fields perhaps buried, possibly become single Am stars.
\citet{irwin-bouvier09} discuss the importance of the pre-main sequence disks
in regulating the rotational period of the stars.
They suggest that the disk holds the angular velocity of the star
by removing angular momentum during the lifetime of the disk (which, in turn,
depends on stellar mass).
For instance, they showed that pre-main sequence stars in the Orion Nebula Cluster (ONC)
with no detectable disk, rotate faster than stars with disks \citep{herbst01,irwin-bouvier09}.
This shows that the disk could also play a role in determining the rotation of the star.
We wonder if single Am stars could perhaps present long-lived pre-main sequence disks,
or present an efficient removal of angular momentum through jets and outflows,
resulting in the slow rotation of Am stars.
These speculative scenarios would require further investigation.

We also note that \citet{abt79} found likely Am stars in young clusters
with rotational velocities (v $\sin i$) greater than $\sim$120 km/s.
This would imply that Am stars are not necessarily slow rotators.
However, diffusion models of Am stars require slow rotation, and the result of \citet{abt79}
is based on few data points \citep{murphy13,stepien00}.
It would be valuable to check the result of \citet{abt79} about the possible
high rotation for some Am stars.

\section{Conclusions}

We performed a detailed abundance analysis of a sample of 19 early-type stars, 7 of them hosting
hot-BDs and 12 of them hot-Jupiter planets, in order to determine if the presence of these hot-low mass
companions could have an effect on the chemical pattern of their host stars.
In particular, we wanted to determine if hot-low mass companions could play a role
in the possible development of Am stars. The main results of this work are as follows:

-The stars studied do not show a single abundance pattern. Instead, the stars
analyzed show near-solar abundance values, slightly subsolar and also suprasolar abundances,
including the detection of four Am stars (KELT-19A, KELT-17, HATS-70 and TOI-503)
and two possible Am stars (TOI-681 and HAT-P-69).

-We detected the new Am star HATS-70 which hosts a hot-BD, and rule out
this class for the hot-Jupiter host WASP-189, different than previous reports.
This highlights the importance to perform a detailed chemical analysis.

-We obtained a rough estimation for the incidence of Am stars within stars
hosting hot-BDs (50-75\%) and within stars hosting hot-Jupiters (20-42\%).
In particular, the incidence of Am stars hosting hot-BDs resulted higher than
the frequency of Am stars in general (31-36\%).
These starting trends would indicate that the presence of hot-BDs could possibly
play a role in the development of Am stars, different to the case of hot-Jupiter planets.
This is possibly due to the higher mass of BDs compared to the giant planets, 
or to a different formation mechanism.
In this way, hot-BDs could possibly help to mitigate the "single Am" problem.

-If confirmed, the initial trends found here would indicate that the search of hot-BDs
could be benefited by targeting single Am stars or Am stars in wide binary systems.

The detection of additional low-mass companions around early-type stars
such as those obtained from the survey KELT, the satellite TESS and the future mission PLATO
would help to increase the significance of the interesting results found here.

\begin{acknowledgements}
We thank the anonymous referee for constructive comments that greatly improved the paper.
The authors thank Dr. R. Kurucz for making their codes available to us.
CS acknowledge financial support from FONCyT (Argentina) through grant PICT 2017-2294
and the National University of San Juan (Argentina) through grant CICITCA E1134.
PM and JA acknowledge financial support from CONICET (Argentina).
IRAF is distributed by the National Optical Astronomical Observatories, 
which is operated by the Association of Universities for Research in Astronomy, Inc., under a cooperative agreement
with the National Science Foundation.
Based on data acquired at Complejo Astron\'omico El Leoncito, operated under agreement between
the Consejo Nacional de Investigaciones Cient\'ificas y T\'ecnicas de la Rep\'ublica Argentina and
the National Universities of La Plata, C\'ordoba and San Juan.

This paper includes data collected by the TESS mission that are publicly available from the Mikulski Archive for Space Telescopes (MAST).
We acknowledge the use of public TESS data from pipelines at the TESS Science Office and at the TESS Science Processing Operations Center.
Resources supporting this work were provided by the NASA High-End Computing (HEC) Program through the NASA Advanced Supercomputing (NAS) Division
at Ames Research Center for the production of the SPOC data products. Funding for the TESS mission is provided by NASA's Science Mission Directorate.

This study is based on observations obtained through the Gemini Remote Access to CFHT ESPaDOnS Spectrograph (GRACES).
ESPaDOnS is located at the Canada–France–Hawaii Telescope (CFHT), which is operated by the National Research Council of
Canada, the Institut National des Sciences de l Univers of the Centre National de la Recherche Scientifique of France, and the
University of Hawaii. ESPaDOnS is a collaborative project funded by France (CNRS, MENESR, OMP, LATT), Canada (NSERC),
CFHT, and ESA. ESPaDOnS was remotely controlled from the international Gemini Observatory, a program of NSFs NOIRLab,
which is managed by the Association of Universities for Research in Astronomy (AURA) under a cooperative agreement with the
National Science Foundation on behalf of the Gemini partnership: the National Science Foundation (United States), the National
Research Council (Canada), Agencia Nacional de Investigación y Desarrollo (Chile), Ministerio de Ciencia, Tecnología e Innovación
(Argentina), Ministério da Ciência, Tecnologia e Inovação (Brazil), and Korea Astronomy and Space Science Institute (Republic of
Korea).

Based on observations collected with the SOPHIE spectrograph on the 1.93 m telescope at the Observatoire de Haute-Provence (CNRS), France, by the SOPHIE Consortium.

\end{acknowledgements}

\begin{appendix}

\FloatBarrier
\clearpage

\section{Chemical abundances}

We present in this section the chemical abundances obtained for the stars in this work.
The columns in the Tables \ref{tab.abunds.CoRoT-3} to \ref{tab.abunds.HAT-P-69} 
present the chemical specie, average abundance $\pm$ total error e$_{tot}$, the line-to-line dispersion e$_{1}$,
the error in the abundances (e$_{2}$, e$_{3}$, and e$_{4}$) when varying T$_{\rm eff}$, $\log g$,
and v$_\mathrm{micro}$ by their corresponding uncertainties, and the number of lines n.
For more details, see Section 3.

\begin{table}
\vskip -10.00in
\centering
\caption{Chemical abundances for the star CoRoT-3.}
\begin{tabular}{lrccccc}
\hline
\hline
Specie     & [X/H] $\pm$ e$_{tot}$ & e$_{1}$ & e$_{2}$ & e$_{3}$ & e$_{4}$ & n \\
\hline
\ion{C}{I}    &  0.02 $\pm$ 0.09 & 0.03 & 0.07 & 0.05 & 0.01 & 3 \\ 
\ion{Na}{I}   &  0.25 $\pm$ 0.15 & 0.11 & 0.06 & 0.02 & 0.08 & 1 \\ 
\ion{Mg}{I}   &  0.05 $\pm$ 0.13 & 0.06 & 0.09 & 0.07 & 0.04 & 3 \\ 
\ion{Mg}{II}  &  0.16 $\pm$ 0.19 & 0.11 & 0.09 & 0.04 & 0.12 & 1 \\ 
\ion{Al}{I}   & -0.48 $\pm$ 0.18 & 0.11 & 0.10 & 0.05 & 0.08 & 1 \\ 
\ion{Si}{II}  &  0.01 $\pm$ 0.15 & 0.10 & 0.09 & 0.06 & 0.05 & 2 \\ 
\ion{Ca}{I}   &  0.10 $\pm$ 0.21 & 0.11 & 0.16 & 0.07 & 0.03 & 1 \\ 
\ion{Ca}{II}  &  0.04 $\pm$ 0.13 & 0.11 & 0.04 & 0.06 & 0.03 & 1 \\ 
\ion{Sc}{II}  & -0.09 $\pm$ 0.19 & 0.05 & 0.03 & 0.05 & 0.17 & 3 \\ 
\ion{Ti}{II}  &  0.01 $\pm$ 0.15 & 0.03 & 0.06 & 0.04 & 0.13 & 22 \\ 
\ion{Cr}{II}  &  0.01 $\pm$ 0.10 & 0.03 & 0.02 & 0.04 & 0.09 & 14 \\ 
\ion{Mn}{I}   &  0.11 $\pm$ 0.14 & 0.05 & 0.08 & 0.01 & 0.10 & 10 \\ 
\ion{Fe}{I}   &  0.01 $\pm$ 0.17 & 0.01 & 0.10 & 0.03 & 0.13 & 67 \\ 
\ion{Fe}{II}  &  0.01 $\pm$ 0.14 & 0.02 & 0.07 & 0.01 & 0.12 & 26 \\ 
\ion{Co}{I}   &  0.09 $\pm$ 0.23 & 0.11 & 0.17 & 0.02 & 0.11 & 1 \\ 
\ion{Zn}{I}   & -0.18 $\pm$ 0.12 & 0.04 & 0.07 & 0.01 & 0.08 & 2 \\ 
\ion{Sr}{II}  &  0.29 $\pm$ 0.14 & 0.01 & 0.12 & 0.01 & 0.08 & 2 \\ 
\ion{Y}{II}   &  0.05 $\pm$ 0.09 & 0.01 & 0.03 & 0.05 & 0.07 & 3 \\ 
\ion{Zr}{II}  &  0.18 $\pm$ 0.07 & 0.03 & 0.01 & 0.06 & 0.03 & 2 \\ 
\ion{Ba}{II}  &  0.37 $\pm$ 0.26 & 0.14 & 0.09 & 0.02 & 0.20 & 3 \\ 
\hline
\end{tabular}
\label{tab.abunds.CoRoT-3}
\end{table}

\begin{table}
\centering
\caption{Chemical abundances for the star KELT-1.}
\begin{tabular}{lrccccc}
\hline
\hline
Specie     & [X/H] $\pm$ e$_{tot}$ & e$_{1}$ & e$_{2}$ & e$_{3}$ & e$_{4}$ & n \\
\hline
\ion{C}{I}    &  0.03 $\pm$ 0.12 & 0.01 & 0.03 & 0.09 & 0.08 & 2 \\ 
\ion{Na}{I}   &  0.41 $\pm$ 0.12 & 0.04 & 0.08 & 0.05 & 0.06 & 2 \\ 
\ion{Mg}{I}   &  0.16 $\pm$ 0.18 & 0.06 & 0.13 & 0.10 & 0.03 & 3 \\ 
\ion{Al}{I}   & -0.17 $\pm$ 0.21 & 0.12 & 0.15 & 0.09 & 0.03 & 1 \\ 
\ion{Si}{II}  &  0.26 $\pm$ 0.23 & 0.12 & 0.18 & 0.06 & 0.03 & 1 \\ 
\ion{Ca}{I}   &  0.20 $\pm$ 0.29 & 0.12 & 0.23 & 0.11 & 0.05 & 1 \\ 
\ion{Ca}{II}  &  0.37 $\pm$ 0.22 & 0.12 & 0.14 & 0.11 & 0.01 & 1 \\ 
\ion{Sc}{II}  & -0.13 $\pm$ 0.17 & 0.12 & 0.01 & 0.09 & 0.07 & 1 \\ 
\ion{Ti}{II}  &  0.09 $\pm$ 0.13 & 0.06 & 0.01 & 0.06 & 0.10 & 11 \\ 
\ion{Cr}{II}  &  0.13 $\pm$ 0.08 & 0.05 & 0.01 & 0.04 & 0.05 & 12 \\ 
\ion{Mn}{I}   &  0.23 $\pm$ 0.16 & 0.04 & 0.09 & 0.03 & 0.12 & 4 \\ 
\ion{Fe}{I}   &  0.16 $\pm$ 0.22 & 0.02 & 0.17 & 0.06 & 0.13 & 51 \\ 
\ion{Fe}{II}  &  0.15 $\pm$ 0.18 & 0.02 & 0.12 & 0.02 & 0.12 & 24 \\ 
\ion{Zn}{I}   & -0.26 $\pm$ 0.13 & 0.04 & 0.12 & 0.01 & 0.04 & 2 \\ 
\ion{Sr}{II}  &  0.29 $\pm$ 0.20 & 0.04 & 0.19 & 0.01 & 0.07 & 2 \\ 
\ion{Y}{II}   &  0.06 $\pm$ 0.23 & 0.14 & 0.11 & 0.11 & 0.10 & 2 \\ 
\ion{Ba}{II}  &  0.33 $\pm$ 0.16 & 0.04 & 0.07 & 0.04 & 0.14 & 2 \\ 
\hline
\end{tabular}
\label{tab.abunds.KELT-1}
\end{table}

\begin{table}
\centering
\caption{Chemical abundances for the star HATS-70.}
\begin{tabular}{lrccccc}
\hline
\hline
Specie     & [X/H] $\pm$ e$_{tot}$ & e$_{1}$ & e$_{2}$ & e$_{3}$ & e$_{4}$ & n \\
\hline
\ion{O}{I}    &  0.70 $\pm$ 0.22 & 0.03 & 0.02 & 0.01 & 0.22 & 3 \\ 
\ion{Na}{I}   &  0.09 $\pm$ 0.15 & 0.13 & 0.05 & 0.01 & 0.04 & 1 \\ 
\ion{Mg}{I}   & -0.09 $\pm$ 0.23 & 0.16 & 0.06 & 0.02 & 0.16 & 3 \\ 
\ion{Mg}{II}  &  0.03 $\pm$ 0.19 & 0.13 & 0.09 & 0.04 & 0.10 & 1 \\ 
\ion{Al}{I}   &  0.03 $\pm$ 0.23 & 0.13 & 0.12 & 0.01 & 0.14 & 1 \\ 
\ion{Si}{II}  &  0.25 $\pm$ 0.14 & 0.08 & 0.09 & 0.04 & 0.06 & 4 \\ 
\ion{Ca}{I}   & -0.37 $\pm$ 0.25 & 0.13 & 0.15 & 0.01 & 0.14 & 1 \\ 
\ion{Ca}{II}  & -0.29 $\pm$ 0.15 & 0.13 & 0.06 & 0.02 & 0.01 & 1 \\ 
\ion{Sc}{II}  & -0.56 $\pm$ 0.18 & 0.13 & 0.04 & 0.05 & 0.11 & 1 \\ 
\ion{Ti}{II}  &  0.42 $\pm$ 0.18 & 0.04 & 0.03 & 0.04 & 0.16 & 24 \\ 
\ion{Cr}{II}  &  0.51 $\pm$ 0.14 & 0.05 & 0.01 & 0.04 & 0.12 & 18 \\ 
\ion{Mn}{I}   & -0.13 $\pm$ 0.15 & 0.10 & 0.07 & 0.01 & 0.08 & 4 \\ 
\ion{Fe}{I}   &  0.18 $\pm$ 0.10 & 0.02 & 0.07 & 0.01 & 0.07 & 34 \\ 
\ion{Fe}{II}  &  0.22 $\pm$ 0.11 & 0.04 & 0.05 & 0.02 & 0.09 & 29 \\ 
\ion{Ni}{II}  &  0.50 $\pm$ 0.14 & 0.13 & 0.03 & 0.02 & 0.02 & 1 \\ 
\ion{Zn}{I}   &  0.50 $\pm$ 0.09 & 0.04 & 0.07 & 0.01 & 0.03 & 3 \\ 
\ion{Sr}{II}  &  1.08 $\pm$ 0.22 & 0.12 & 0.11 & 0.02 & 0.15 & 2 \\ 
\ion{Y}{II}   &  1.02 $\pm$ 0.14 & 0.08 & 0.03 & 0.04 & 0.11 & 5 \\ 
\ion{Zr}{II}  &  0.65 $\pm$ 0.18 & 0.17 & 0.03 & 0.04 & 0.01 & 3 \\ 
\ion{Ba}{II}  &  1.90 $\pm$ 0.28 & 0.17 & 0.11 & 0.01 & 0.19 & 4 \\ 
\ion{Nd}{II}  &  1.17 $\pm$ 0.13 & 0.07 & 0.10 & 0.03 & 0.01 & 3 \\ 
\hline
\end{tabular}
\label{tab.abunds.HATS-70}
\end{table}

\begin{table}
\centering
\caption{Chemical abundances for the star MASCARA-4.}
\begin{tabular}{lrccccc}
\hline
\hline
Specie     & [X/H] $\pm$ e$_{tot}$ & e$_{1}$ & e$_{2}$ & e$_{3}$ & e$_{4}$ & n \\
\hline
\ion{C}{I}    & -0.33 $\pm$ 0.04 & 0.03 & 0.01 & 0.02 & 0.01 & 4 \\ 
\ion{O}{I}    &  0.04 $\pm$ 0.25 & 0.03 & 0.02 & 0.01 & 0.24 & 3 \\ 
\ion{Na}{I}   & -0.13 $\pm$ 0.08 & 0.07 & 0.04 & 0.01 & 0.03 & 2 \\ 
\ion{Mg}{I}   & -0.30 $\pm$ 0.23 & 0.09 & 0.04 & 0.02 & 0.21 & 3 \\ 
\ion{Mg}{II}  & -0.10 $\pm$ 0.18 & 0.10 & 0.09 & 0.04 & 0.11 & 1 \\ 
\ion{Al}{I}   & -0.64 $\pm$ 0.29 & 0.25 & 0.06 & 0.01 & 0.14 & 2 \\ 
\ion{Si}{II}  & -0.02 $\pm$ 0.13 & 0.09 & 0.08 & 0.04 & 0.04 & 4 \\ 
\ion{Ca}{I}   & -0.31 $\pm$ 0.22 & 0.10 & 0.12 & 0.01 & 0.15 & 1 \\ 
\ion{Sc}{II}  & -0.26 $\pm$ 0.20 & 0.01 & 0.03 & 0.04 & 0.19 & 2 \\ 
\ion{Ti}{II}  & -0.30 $\pm$ 0.14 & 0.03 & 0.03 & 0.04 & 0.13 & 27 \\ 
\ion{Cr}{II}  & -0.17 $\pm$ 0.06 & 0.02 & 0.01 & 0.03 & 0.05 & 21 \\ 
\ion{Mn}{I}   & -0.26 $\pm$ 0.13 & 0.07 & 0.08 & 0.01 & 0.08 & 8 \\ 
\ion{Fe}{I}   & -0.26 $\pm$ 0.16 & 0.01 & 0.07 & 0.01 & 0.14 & 67 \\ 
\ion{Fe}{II}  & -0.17 $\pm$ 0.12 & 0.02 & 0.04 & 0.02 & 0.11 & 37 \\ 
\ion{Ni}{II}  &  0.11 $\pm$ 0.10 & 0.10 & 0.01 & 0.02 & 0.01 & 1 \\ 
\ion{Zn}{I}   & -0.02 $\pm$ 0.09 & 0.07 & 0.06 & 0.01 & 0.02 & 2 \\ 
\ion{Sr}{II}  &  0.42 $\pm$ 0.20 & 0.08 & 0.09 & 0.04 & 0.16 & 2 \\ 
\ion{Y}{II}   &  0.34 $\pm$ 0.08 & 0.05 & 0.04 & 0.03 & 0.04 & 7 \\ 
\ion{Zr}{II}  &  0.39 $\pm$ 0.11 & 0.09 & 0.03 & 0.04 & 0.02 & 5 \\ 
\ion{Ba}{II}  &  0.53 $\pm$ 0.27 & 0.10 & 0.10 & 0.02 & 0.23 & 4 \\ 
\ion{Eu}{II}  &  0.64 $\pm$ 0.14 & 0.10 & 0.07 & 0.07 & 0.01 & 1 \\ 
\hline
\end{tabular}
\label{tab.abunds.MASCARA-4}
\end{table}

\begin{table}
\centering
\caption{Chemical abundances for the star KELT-25.}
\begin{tabular}{lrccccc}
\hline
\hline
Specie     & [X/H] $\pm$ e$_{tot}$ & e$_{1}$ & e$_{2}$ & e$_{3}$ & e$_{4}$ & n \\
\hline
\ion{C}{I}    & -0.11 $\pm$ 0.12 & 0.11 & 0.05 & 0.02 & 0.01 & 3 \\ 
\ion{O}{I}    &  0.46 $\pm$ 0.22 & 0.07 & 0.01 & 0.01 & 0.21 & 3 \\ 
\ion{Mg}{I}   & -0.09 $\pm$ 0.22 & 0.09 & 0.10 & 0.04 & 0.17 & 3 \\ 
\ion{Mg}{II}  &  0.02 $\pm$ 0.14 & 0.13 & 0.06 & 0.04 & 0.01 & 2 \\ 
\ion{Al}{I}   & -0.05 $\pm$ 0.21 & 0.18 & 0.05 & 0.04 & 0.09 & 2 \\ 
\ion{Si}{II}  &  0.06 $\pm$ 0.20 & 0.17 & 0.08 & 0.05 & 0.01 & 3 \\ 
\ion{Ca}{I}   & -0.23 $\pm$ 0.29 & 0.14 & 0.16 & 0.02 & 0.19 & 1 \\ 
\ion{Ca}{II}  & -0.05 $\pm$ 0.14 & 0.01 & 0.12 & 0.01 & 0.07 & 2 \\ 
\ion{Sc}{II}  & -0.18 $\pm$ 0.20 & 0.02 & 0.08 & 0.06 & 0.17 & 2 \\ 
\ion{Ti}{II}  &  0.03 $\pm$ 0.19 & 0.06 & 0.04 & 0.06 & 0.17 & 13 \\ 
\ion{Cr}{II}  &  0.03 $\pm$ 0.12 & 0.06 & 0.03 & 0.05 & 0.09 & 10 \\ 
\ion{Mn}{I}   & -0.18 $\pm$ 0.21 & 0.17 & 0.12 & 0.01 & 0.04 & 2 \\ 
\ion{Fe}{I}   & -0.14 $\pm$ 0.17 & 0.02 & 0.12 & 0.01 & 0.12 & 31 \\ 
\ion{Fe}{II}  & -0.05 $\pm$ 0.15 & 0.03 & 0.08 & 0.03 & 0.12 & 23 \\ 
\ion{Sr}{II}  & -0.18 $\pm$ 0.24 & 0.09 & 0.12 & 0.09 & 0.16 & 2 \\ 
\ion{Y}{II}   &  0.23 $\pm$ 0.11 & 0.06 & 0.04 & 0.07 & 0.04 & 3 \\ 
\ion{Zr}{II}  &  0.15 $\pm$ 0.19 & 0.14 & 0.12 & 0.04 & 0.02 & 1 \\ 
\ion{Ba}{II}  &  0.37 $\pm$ 0.24 & 0.14 & 0.12 & 0.04 & 0.15 & 2 \\ 
\hline
\end{tabular}
\label{tab.abunds.KELT-25}
\end{table}

\begin{table}
\centering
\caption{Chemical abundances for the star HAT-P-70.}
\begin{tabular}{lrccccc}
\hline
\hline
Specie     & [X/H] $\pm$ e$_{tot}$ & e$_{1}$ & e$_{2}$ & e$_{3}$ & e$_{4}$ & n \\
\hline
\ion{C}{I}    & -0.34 $\pm$ 0.12 & 0.09 & 0.07 & 0.02 & 0.01 & 2 \\ 
\ion{O}{I}    &  0.46 $\pm$ 0.20 & 0.05 & 0.04 & 0.01 & 0.19 & 3 \\ 
\ion{Mg}{I}   & -0.33 $\pm$ 0.23 & 0.10 & 0.13 & 0.04 & 0.16 & 3 \\ 
\ion{Al}{I}   & -0.56 $\pm$ 0.22 & 0.17 & 0.10 & 0.01 & 0.11 & 2 \\ 
\ion{Si}{II}  & -0.26 $\pm$ 0.22 & 0.12 & 0.10 & 0.05 & 0.15 & 3 \\ 
\ion{Ca}{I}   & -0.35 $\pm$ 0.26 & 0.14 & 0.19 & 0.01 & 0.12 & 1 \\ 
\ion{Ca}{II}  & -0.24 $\pm$ 0.17 & 0.06 & 0.15 & 0.01 & 0.07 & 2 \\ 
\ion{Sc}{II}  & -0.44 $\pm$ 0.15 & 0.04 & 0.10 & 0.04 & 0.09 & 2 \\ 
\ion{Ti}{II}  & -0.10 $\pm$ 0.17 & 0.04 & 0.08 & 0.05 & 0.13 & 18 \\ 
\ion{Cr}{II}  & -0.05 $\pm$ 0.15 & 0.10 & 0.03 & 0.03 & 0.09 & 6 \\ 
\ion{Fe}{I}   & -0.39 $\pm$ 0.19 & 0.02 & 0.15 & 0.01 & 0.11 & 38 \\ 
\ion{Fe}{II}  & -0.29 $\pm$ 0.15 & 0.03 & 0.11 & 0.02 & 0.10 & 28 \\ 
\ion{Sr}{II}  & -0.88 $\pm$ 0.26 & 0.19 & 0.11 & 0.04 & 0.13 & 2 \\ 
\ion{Zr}{II}  & -0.03 $\pm$ 0.19 & 0.14 & 0.12 & 0.02 & 0.01 & 1 \\ 
\ion{Ba}{II}  & -0.04 $\pm$ 0.24 & 0.14 & 0.19 & 0.01 & 0.04 & 1 \\ 
\hline
\end{tabular}
\label{tab.abunds.HAT-P-70}
\end{table}

\begin{table}
\centering
\caption{Chemical abundances for the star TOI-503.}
\begin{tabular}{lrccccc}
\hline
\hline
Specie     & [X/H] $\pm$ e$_{tot}$ & e$_{1}$ & e$_{2}$ & e$_{3}$ & e$_{4}$ & n \\
\hline
\ion{Li}{I}   &  1.89 $\pm$ 0.11 & 0.09 & 0.06 & 0.01 & 0.01 & 1 \\ 
\ion{C}{I}    & -0.70 $\pm$ 0.09 & 0.09 & 0.01 & 0.02 & 0.01 & 1 \\ 
\ion{O}{I}    & -0.45 $\pm$ 0.19 & 0.01 & 0.03 & 0.02 & 0.19 & 3 \\ 
\ion{Na}{I}   &  0.41 $\pm$ 0.08 & 0.04 & 0.04 & 0.01 & 0.06 & 2 \\ 
\ion{Mg}{I}   & -0.14 $\pm$ 0.18 & 0.11 & 0.05 & 0.01 & 0.13 & 2 \\ 
\ion{Mg}{II}  & -0.28 $\pm$ 0.17 & 0.09 & 0.09 & 0.04 & 0.11 & 1 \\ 
\ion{Si}{II}  &  0.16 $\pm$ 0.14 & 0.09 & 0.08 & 0.04 & 0.06 & 1 \\ 
\ion{Ca}{II}  & -0.45 $\pm$ 0.21 & 0.07 & 0.09 & 0.01 & 0.19 & 2 \\ 
\ion{Sc}{II}  & -0.49 $\pm$ 0.10 & 0.09 & 0.02 & 0.03 & 0.01 & 1 \\ 
\ion{Ti}{II}  &  0.05 $\pm$ 0.15 & 0.05 & 0.02 & 0.03 & 0.14 & 14 \\ 
\ion{Cr}{II}  &  0.46 $\pm$ 0.11 & 0.02 & 0.01 & 0.03 & 0.10 & 19 \\ 
\ion{Mn}{I}   &  0.23 $\pm$ 0.08 & 0.05 & 0.05 & 0.01 & 0.04 & 3 \\ 
\ion{Fe}{I}   &  0.21 $\pm$ 0.19 & 0.02 & 0.05 & 0.01 & 0.18 & 34 \\ 
\ion{Fe}{II}  &  0.23 $\pm$ 0.13 & 0.02 & 0.03 & 0.02 & 0.13 & 29 \\ 
\ion{Zn}{I}   &  0.43 $\pm$ 0.08 & 0.02 & 0.05 & 0.01 & 0.06 & 2 \\ 
\ion{Sr}{II}  &  0.20 $\pm$ 0.23 & 0.09 & 0.09 & 0.07 & 0.18 & 1 \\ 
\ion{Y}{II}   &  0.57 $\pm$ 0.10 & 0.08 & 0.03 & 0.03 & 0.03 & 4 \\ 
\ion{Ba}{II}  &  1.33 $\pm$ 0.16 & 0.08 & 0.08 & 0.01 & 0.11 & 4 \\ 
\hline
\end{tabular}
\label{tab.abunds.TOI-503}
\end{table}

\begin{table}
\centering
\caption{Chemical abundances for the star TOI-681.}
\begin{tabular}{lrccccc}
\hline
\hline
Specie     & [X/H] $\pm$ e$_{tot}$ & e$_{1}$ & e$_{2}$ & e$_{3}$ & e$_{4}$ & n \\
\hline
\ion{C}{I}    & -0.05 $\pm$ 0.13 & 0.13 & 0.01 & 0.02 & 0.01 & 1 \\ 
\ion{Na}{I}   &  0.51 $\pm$ 0.12 & 0.07 & 0.05 & 0.02 & 0.08 & 2 \\ 
\ion{Mg}{I}   & -0.29 $\pm$ 0.19 & 0.05 & 0.08 & 0.04 & 0.15 & 3 \\ 
\ion{Mg}{II}  & -0.20 $\pm$ 0.19 & 0.13 & 0.07 & 0.02 & 0.12 & 1 \\ 
\ion{Si}{II}  &  0.42 $\pm$ 0.17 & 0.13 & 0.09 & 0.04 & 0.05 & 1 \\ 
\ion{Ca}{I}   & -0.37 $\pm$ 0.28 & 0.13 & 0.20 & 0.04 & 0.15 & 1 \\ 
\ion{Sc}{II}  &  0.13 $\pm$ 0.15 & 0.01 & 0.08 & 0.02 & 0.13 & 3 \\ 
\ion{Ti}{II}  &  0.23 $\pm$ 0.18 & 0.05 & 0.05 & 0.04 & 0.16 & 15 \\ 
\ion{Cr}{II}  &  0.35 $\pm$ 0.08 & 0.03 & 0.01 & 0.03 & 0.06 & 14 \\ 
\ion{Mn}{I}   &  0.02 $\pm$ 0.12 & 0.07 & 0.07 & 0.03 & 0.06 & 4 \\ 
\ion{Fe}{I}   &  0.05 $\pm$ 0.20 & 0.02 & 0.11 & 0.02 & 0.17 & 35 \\ 
\ion{Fe}{II}  &  0.07 $\pm$ 0.14 & 0.03 & 0.07 & 0.01 & 0.12 & 15 \\ 
\ion{Zn}{I}   &  0.20 $\pm$ 0.10 & 0.01 & 0.09 & 0.03 & 0.05 & 2 \\ 
\ion{Sr}{II}  &  0.33 $\pm$ 0.27 & 0.19 & 0.15 & 0.03 & 0.13 & 2 \\ 
\ion{Y}{II}   &  0.30 $\pm$ 0.07 & 0.04 & 0.01 & 0.05 & 0.03 & 3 \\ 
\ion{Ba}{II}  &  1.13 $\pm$ 0.18 & 0.10 & 0.11 & 0.05 & 0.09 & 3 \\ 
\hline
\end{tabular}
\label{tab.abunds.TOI-681}
\end{table}

\begin{table}
\centering
\caption{Chemical abundances for the star TOI-587.}
\begin{tabular}{lrccccc}
\hline
\hline
Specie     & [X/H] $\pm$ e$_{tot}$ & e$_{1}$ & e$_{2}$ & e$_{3}$ & e$_{4}$ & n \\
\hline
\ion{O}{I}    & -0.38 $\pm$ 0.05 & 0.01 & 0.04 & 0.02 & 0.01 & 2 \\ 
\ion{Mg}{I}   & -0.48 $\pm$ 0.24 & 0.09 & 0.21 & 0.07 & 0.02 & 3 \\ 
\ion{Mg}{II}  & -0.58 $\pm$ 0.09 & 0.09 & 0.01 & 0.01 & 0.01 & 3 \\ 
\ion{Al}{II}  &  0.07 $\pm$ 0.11 & 0.08 & 0.06 & 0.05 & 0.01 & 1 \\ 
\ion{Si}{II}  & -0.20 $\pm$ 0.06 & 0.04 & 0.05 & 0.01 & 0.02 & 4 \\ 
\ion{Ca}{I}   & -0.72 $\pm$ 0.20 & 0.08 & 0.17 & 0.06 & 0.01 & 1 \\ 
\ion{Sc}{II}  & -0.77 $\pm$ 0.23 & 0.21 & 0.10 & 0.01 & 0.01 & 2 \\ 
\ion{Ti}{II}  & -0.63 $\pm$ 0.14 & 0.02 & 0.13 & 0.01 & 0.02 & 12 \\ 
\ion{Cr}{II}  & -0.22 $\pm$ 0.08 & 0.05 & 0.05 & 0.02 & 0.01 & 11 \\ 
\ion{Fe}{I}   & -0.16 $\pm$ 0.17 & 0.02 & 0.17 & 0.02 & 0.03 & 18 \\ 
\ion{Fe}{II}  & -0.13 $\pm$ 0.07 & 0.01 & 0.05 & 0.02 & 0.04 & 29 \\ 
\ion{Ni}{II}  &  0.21 $\pm$ 0.10 & 0.08 & 0.05 & 0.02 & 0.03 & 1 \\ 
\ion{Sr}{II}  &  0.31 $\pm$ 0.24 & 0.14 & 0.13 & 0.02 & 0.15 & 2 \\ 
\ion{Ba}{II}  &  0.87 $\pm$ 0.29 & 0.18 & 0.23 & 0.04 & 0.01 & 2 \\ 
\hline
\end{tabular}
\label{tab.abunds.TOI-587}
\end{table}

\begin{table}
\centering
\caption{Chemical abundances for the star KELT-19A.}
\begin{tabular}{lrccccc}
\hline
\hline
Specie     & [X/H] $\pm$ e$_{tot}$ & e$_{1}$ & e$_{2}$ & e$_{3}$ & e$_{4}$ & n \\
\hline
\ion{O}{I}    & -0.57 $\pm$ 0.25 & 0.20 & 0.04 & 0.05 & 0.14 & 1 \\ 
\ion{Na}{I}   &  0.51 $\pm$ 0.22 & 0.20 & 0.01 & 0.01 & 0.10 & 1 \\ 
\ion{Mg}{I}   & -0.10 $\pm$ 0.18 & 0.02 & 0.04 & 0.08 & 0.15 & 3 \\ 
\ion{Mg}{II}  & -0.18 $\pm$ 0.24 & 0.20 & 0.05 & 0.03 & 0.13 & 1 \\ 
\ion{Si}{II}  &  0.37 $\pm$ 0.22 & 0.20 & 0.07 & 0.05 & 0.04 & 1 \\ 
\ion{Ca}{II}  & -0.55 $\pm$ 0.18 & 0.07 & 0.06 & 0.01 & 0.15 & 2 \\ 
\ion{Sc}{II}  & -0.79 $\pm$ 0.16 & 0.09 & 0.01 & 0.07 & 0.10 & 2 \\ 
\ion{Ti}{II}  &  0.39 $\pm$ 0.18 & 0.05 & 0.02 & 0.06 & 0.16 & 9 \\ 
\ion{Cr}{II}  &  0.62 $\pm$ 0.11 & 0.03 & 0.01 & 0.03 & 0.10 & 13 \\ 
\ion{Mn}{I}   &  0.43 $\pm$ 0.21 & 0.20 & 0.03 & 0.02 & 0.05 & 1 \\ 
\ion{Fe}{I}   &  0.39 $\pm$ 0.22 & 0.04 & 0.06 & 0.03 & 0.21 & 20 \\ 
\ion{Fe}{II}  &  0.41 $\pm$ 0.17 & 0.04 & 0.02 & 0.01 & 0.16 & 14 \\ 
\ion{Zn}{I}   &  0.34 $\pm$ 0.11 & 0.09 & 0.04 & 0.01 & 0.05 & 2 \\ 
\ion{Sr}{II}  &  0.58 $\pm$ 0.28 & 0.20 & 0.11 & 0.05 & 0.15 & 1 \\ 
\ion{Y}{II}   &  1.23 $\pm$ 0.26 & 0.12 & 0.03 & 0.14 & 0.18 & 3 \\ 
\ion{Ba}{II}  &  1.46 $\pm$ 0.19 & 0.11 & 0.06 & 0.03 & 0.14 & 2 \\ 
\hline
\end{tabular}
\label{tab.abunds.KELT-19A}
\end{table}

\begin{table}
\centering
\caption{Chemical abundances for the star HAT-P-69.}
\begin{tabular}{lrccccc}
\hline
\hline
Specie     & [X/H] $\pm$ e$_{tot}$ & e$_{1}$ & e$_{2}$ & e$_{3}$ & e$_{4}$ & n \\
\hline
\ion{O}{I}    & -0.40 $\pm$ 0.18 & 0.04 & 0.04 & 0.04 & 0.17 & 2 \\ 
\ion{Na}{I}   &  0.36 $\pm$ 0.09 & 0.03 & 0.06 & 0.02 & 0.05 & 2 \\ 
\ion{Mg}{I}   & -0.11 $\pm$ 0.21 & 0.09 & 0.09 & 0.01 & 0.16 & 3 \\ 
\ion{Mg}{II}  & -0.15 $\pm$ 0.17 & 0.12 & 0.04 & 0.04 & 0.11 & 1 \\ 
\ion{Si}{II}  &  0.22 $\pm$ 0.19 & 0.12 & 0.14 & 0.02 & 0.05 & 1 \\ 
\ion{Ca}{II}  & -0.05 $\pm$ 0.22 & 0.12 & 0.15 & 0.01 & 0.12 & 2 \\ 
\ion{Sc}{II}  & -0.45 $\pm$ 0.13 & 0.12 & 0.02 & 0.03 & 0.01 & 1 \\ 
\ion{Ti}{II}  &  0.06 $\pm$ 0.15 & 0.04 & 0.01 & 0.03 & 0.14 & 11 \\ 
\ion{Cr}{II}  &  0.17 $\pm$ 0.11 & 0.02 & 0.04 & 0.02 & 0.10 & 9 \\ 
\ion{Mn}{I}   &  0.21 $\pm$ 0.09 & 0.07 & 0.06 & 0.01 & 0.01 & 2 \\ 
\ion{Fe}{I}   &  0.06 $\pm$ 0.22 & 0.02 & 0.10 & 0.01 & 0.19 & 27 \\ 
\ion{Fe}{II}  &  0.07 $\pm$ 0.15 & 0.03 & 0.05 & 0.03 & 0.14 & 23 \\ 
\ion{Zn}{I}   &  0.00 $\pm$ 0.15 & 0.12 & 0.08 & 0.01 & 0.01 & 1 \\ 
\ion{Sr}{II}  &  0.71 $\pm$ 0.24 & 0.12 & 0.09 & 0.04 & 0.19 & 1 \\ 
\ion{Y}{II}   &  0.91 $\pm$ 0.20 & 0.03 & 0.07 & 0.06 & 0.17 & 3 \\ 
\ion{Zr}{II}  &  0.75 $\pm$ 0.22 & 0.19 & 0.11 & 0.01 & 0.05 & 3 \\ 
\ion{Ba}{II}  &  0.91 $\pm$ 0.21 & 0.13 & 0.14 & 0.03 & 0.09 & 2 \\ 
\hline
\end{tabular}
\label{tab.abunds.HAT-P-69}
\end{table}

\end{appendix}

\end{document}